\documentclass{jfm}
\usepackage{graphicx}
\usepackage{epstopdf, epsfig}
\usepackage[caption=false]{subfig}
\usepackage{tikz} 
\usetikzlibrary{shapes}
\usepackage{pgfplots}
\usepackage{natbib}
\usepackage{amsmath}
\usepackage{multirow}
\pgfplotsset{compat=newest}
\pgfplotsset{scaled y ticks=false}
\pgfplotsset{scaled x ticks=false}
\usepackage[normalem]{ulem}
\definecolor{mycolor1}{rgb}{0.07843,0.07843,0.07843}% black
\newcommand{\mylab}[3]{\raisebox{#2}[0mm][0mm]{\makebox[0mm][l]{\hspace*{#1}#3}}}
\newcommand{\hsquare}{\raisebox{0.5pt}{\tikz{\node[draw,scale=0.5,regular polygon, regular polygon sides=4,fill=none](){};}}}
\newcommand{\hcircle}{\raisebox{0.5pt}{\tikz{\node[draw,scale=0.5,circle,fill=none](){};}}}
\newcommand{\blackline}{\raisebox{2pt}{\tikz{\draw[-,mycolor1,solid,line width = 0.75pt](0,0) -- (5mm,0);}}}
\newcommand{\blackdotted}{\raisebox{2pt}{\tikz{\draw[-,mycolor1,dotted,line width = 0.75pt](0,0) -- (5mm,0);}}}
\newcommand{\blackdashed}{\raisebox{2pt}{\tikz{\draw[-,mycolor1,dashed,line width = 0.75pt](0,0) -- (5mm,0);}}}
\definecolor{dgreen}{rgb}{0.0,0.5,0.0}

\shorttitle{Turbulent flows over dense filament canopies}
\shortauthor{A. Sharma and R. Garc{\'i}a-Mayoral}

\title{Turbulent flows over dense filament canopies}

\author{Akshath Sharma\aff{1}
       \and Ricardo Garc{\'i}a-Mayoral\aff{1}
  \corresp{\email{r.gmayoral@eng.cam.ac.uk}}}

\affiliation{\aff{1}Department of Engineering, University of
Cambridge, Trumpington Street, Cambridge CB2~1PZ, UK}

\begin{document}

\maketitle

\begin{abstract}
Turbulent flows over dense canopies consisting of rigid filaments of small size are investigated using direct numerical simulations. {The effect of the height and spacing of the canopy elements on the flow is studied. The flow is composed of an element-coherent, dispersive flow and an incoherent flow, which includes contributions from the background turbulence and from the flow arising from the Kelvin--Helmholtz-like, mixing-layer instability typically reported over dense canopies. For the present canopies, with spacings $s^+ \approx 3$--$50$, the background turbulence is essentially precluded from penetrating within the canopy. As the elements are `tall', with height-to-spacing ratios $h/s \gtrsim 1$, the roughness sublayer of the canopy is determined by their spacing, extending to $y \approx 2$--$3s$ above the canopy tips. The dispersive velocity fluctuations are observed to also depend mainly on the spacing, and are small deep within the canopy, where the footprint of the Kelvin--Helmholtz-like instability dominates. The instability is governed by the canopy drag, which sets the shape of the mean velocity profile, and thus the shear length near the canopy tips. For the tall canopies considered here, this drag is governed by the element spacing and width, that is, the planar layout of the canopy. The mixing length, which determines the lengthscale of the instability, is essentially the sum of its height above and below the canopy tips. The former remains roughly the same in wall-units and the latter is linear with $s$ for all the canopies considered. For very small element spacings, $s^+ \lesssim 10$, the elements obstruct the fluctuations and the instability is inhibited. Within the range of $s^+$ of the present canopies, the obstruction decreases with increasing spacing and the signature of the Kelvin--Helmholtz-like rollers intensifies. For sparser canopies, however, the intensification of the instabilities can be expected to cease as the assumption of a spatially homogeneous mean flow would break down. For the present, dense configurations, the canopy depth also has an influence on the development of the instability. For shallow canopies, $h/s \sim 1$, the lack of depth blocks the Kelvin--Helmholtz-like rollers. For deep canopies, $h/s \gtrsim 6$, the rollers do not perceive the bottom wall and the effect of the canopy height on the flow saturates. Some of the effects of the canopy parameters on the instability can be captured by linear analysis.}

\end{abstract}

\begin{keywords}
\end{keywords}

\section{Introduction}\label{sec:intro}

The present work studies flows over dense canopies of filaments of small size. Canopy flows are mainly studied in the context of natural vegetation canopies, but they also encompass engineering flows where the canopy parameters may be very different from those of natural canopies. Many of the key findings from natural canopy studies have been summarised in the reviews by \citet{Finnigan2000}, \citet{Belcher2012} and \citet{Nepf2012}. In engineering applications, filament canopies can, for instance, be used to enhance heat transfer \citep{Fazu1989,Bejan1993} and for energy harvesting \citep{McGarry2011,Elahi2018}. Depending on the geometry and spacing of their elements, canopies can be classified as sparse, dense or transitional \citep{Nepf2012}. Dense canopies typically have small element spacings compared to the lengthscales in the overlying flow, and thus prevent turbulent eddies from penetrating efficiently within the canopy. Sparse canopies, on the other hand, have large element spacings and consequently, turbulent eddies are essentially able to penetrate the full height of the canopy \citep{Poggi2004,Nepf2012,Sharma2018,AS2019}. Transitional, or intermediate, canopies would lie between these two regimes. {In the present study, we assess how the flow within and above dense canopies is affected by canopy parameters, such as the element height and spacing. The canopies considered have spacings $s^+ \approx \mathcal{O}(10)$, which should be small enough to limit the penetration of the overlying turbulence within them. The frontal area density $\lambda_f$ is also a commonly used measure to categorise canopies \citep{Finnigan2000,Poggi2004,Huang2009,Nepf2012}. Canopies with $\lambda \gg 0.1$ are classified as dense, with $\lambda \approx 0.1$ as transitional and with $\lambda \ll 0.1$ as sparse. However, in addition to $\lambda_f$, the lengthscales of the overlying turbulence should also be considered when determining the canopy regime. A given canopy geometry with a fixed $\lambda_f$ may have element spacings much smaller than any overlying turbulent eddy at a particular Reynolds number, thereby not allowing turbulence to penetrate within the canopy. As the Reynolds number is increased, however, the size of these eddies will eventually become comparable to the element spacing, allowing turbulence to penetrate efficiently within the canopy. To assess this effect, we also study canopies with self-similar geometries, which have a fixed $\lambda_f$, but different sizes in friction units.}

{We also place attention on the effect of canopy parameters on the Kelvin--Helmholtz-like, mixing-layer instability characteristic of dense canopy flows \citep{Raupach1996,Finnigan2000,Nepf2012}. This instability originates} from the inflection point in the mean velocity profile at the canopy-tip plane \citep{Raupach1996}. Kelvin--Helmholtz instabilities manifest as spanwise coherent rollers whose streamwise scale is determined by the shear-layer thickness \citep{Michalke1972,Brown1974}. \citet{Ghisalberti2004} noted that, while in free-shear flows the shear-layer thickness, and consequently the instability wavelength, continues to grow downstream, in fully developed canopy flows this thickness is constant and is set by the net canopy drag. Therefore, a fixed instability wavelength is generally associated with dense canopy flows. Several studies have shown that some aspects of this instability can be captured using a mean-flow linear stability analysis \citep{Raupach1996,White2007,Singh2016,Zampogna2016,Luminari2016}. Some studies have also suggested that at the high Reynolds numbers of natural canopy flows these instabilities can be distorted by the ambient turbulence fluctuations and lose their spanwise-coherent nature \citep{Finnigan2009,Bailey2016}. The importance of this instability decreases as the element spacing is increased, and sparse canopies do not exhibit a notable signature \citep{Poggi2004,Pietri2009,Huang2009,AS2019}.

 {Based on the observations of previous studies, we would expect that the effect of increasing the canopy height for a fixed element spacing on the instability and the surrounding flow would eventually saturate. \citet{Ghisalberti2009} and \citet{Nepf2012} proposed that the effective canopy height perceived by the overlying flow would be a function of the canopy shear-layer thickness as it determined the extent to which the Kelvin--Helmholtz-like instabilities penetrated within the canopy. In flows over arrays of cuboidal posts, \citet{Sadique2017} found that the mean-velocity profiles over them became independent of the element heights at large element aspect ratios.} They concluded that the overlying flow only interacted with the region near the element tips, and that the height below this `active' region was dormant, and did not have a significant effect on the overlying flow. For their geometries, \citet{Sadique2017} observed the height of this active region to be related to the element width. A similar observation was also made by \citet{Macdonald2018}, who performed direct numerical simulations (DNSs) of flows over spanwise-aligned bars. They found that the gap between the bars was the relevant lengthscale for the overlying flow, and that increasing the height of the bars beyond a certain height-to-gap ratio did not affect the overlying flow, or cause an increase in the drag they produced.

{In the present work, we conduct a systematic range of DNSs changing the canopy height and spacing separately in order to study their individual effects on the surrounding turbulence and on the Kelvin--Helmholtz-like instability. We also consider canopy geometries with constant $\lambda_f$ for which the height and spacing are changed simultaneously in a fixed proportion.} The canopies consist of rigid, prismatic filaments with small element spacings and large height-to-spacing ratios. {The element spacings considered, $s^+ \approx 3$--$50$, are much smaller than those typical of most natural canopy flows and would, for instance, be representative of flows over engineered canopies such as those mentioned previously in this section. We also assess how models based on linear stability analysis capture some of the effects of the canopy parameters on the Kelvin--Helmholtz-like instability.}

The paper is organised as follows. The numerical methods used for the simulations and the canopy parameters are discussed in \S\ref{sec:methodology}. The results from the DNSs, detailing the effect of the canopy parameters on the overlying turbulence and the Kelvin--Helmholtz-like instabilities are discussed in \S\ref{sec:turb_fluct}. The results from linear stability analysis and a model to capture the instabilities are presented in \S\ref{sec:stability_analysis}. The conclusions are summarised in \S\ref{sec:conclusions}.

\section{Methodology}\label{sec:methodology}

\begin{figure}
        \centering
		\vspace{0mm} \includegraphics[scale=0.38,trim={0mm 0mm 0mm 0mm},clip]{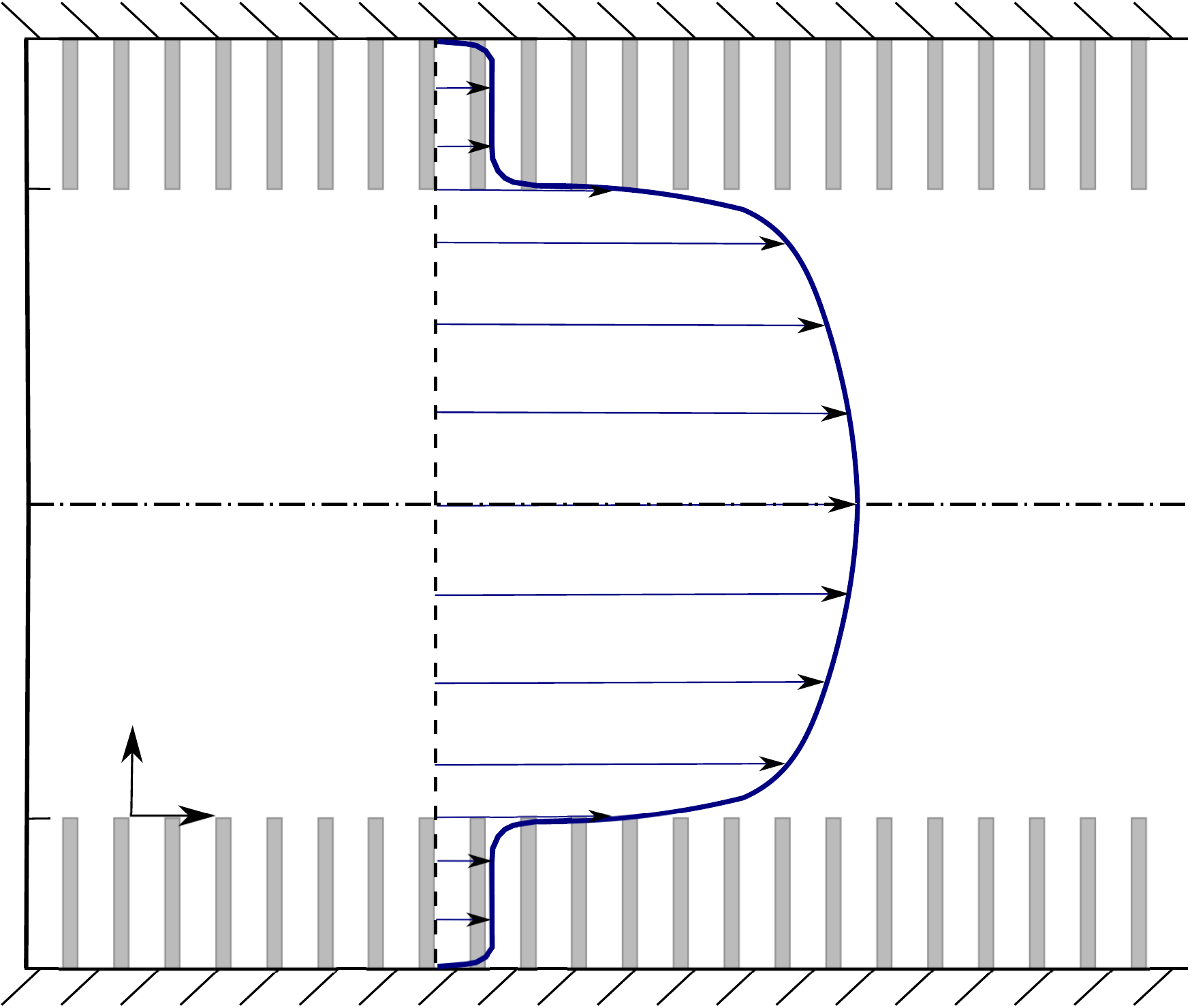}%
  		\mylab{-7.2cm}{0.15cm}{$ -h$}%
  		\mylab{-6.95cm}{1cm}{$ 0$}%
  		\mylab{-6.95cm}{2.8cm}{$\delta$}%
  		 \mylab{-5.7cm}{0.9cm}{$ x$}%
  		 \mylab{-6.3cm}{1.5cm}{$ y$}%
  		 \mylab{-7.1cm}{4.6cm}{$2\delta$}%
  		\mylab{-7.65cm}{5.4cm}{$2\delta+h$}%
  		\caption{Schematic representation of the domain considered in the present study.}%
   	     \label{fig:channel_schem}
\end{figure}

We conduct direct numerical simulations of symmetric channels with rigid canopy elements on both walls. The streamwise, wall-normal and spanwise coordinates are $x$, $y$ and $z$, with the associated velocities $u$, $v$ and $w$, {and $p$ is the kinematic pressure}. The wall-normal origin, $y = 0$, is defined at the tip plane of the canopies protruding from the bottom wall. The channel height, $2\delta$, is defined as the distance between the tip planes of the canopies on the top and bottom walls. The canopy elements, therefore, extend below $y = 0$ and above $y = 2\delta$ and have a height $h$. A schematic representation of the channel is portrayed in figure~\ref{fig:channel_schem}. The size of the domain is a standard $2\pi\delta$ in the streamwise direction and $\pi\delta$ in the spanwise direction. {We use the channel half-height as the length scale in outer units, which implies that $\delta = 1$ in outer scaling}. The domain-to-canopy height ratio for most cases considered here is $(\delta+h)/h \approx 3$. We will show in \S\ref{sec:turb_fluct} that the height of the roughness sublayer scales with the canopy spacing rather than their height, as in the configurations of \citet{Sadique2017} and \citet{Macdonald2018}, and that outer-layer similarity is recovered well below the channel half-height. The channel height to element spacing ratio for most canopies considered is $\delta/s \gtrsim 10$, and for the canopy with the largest element spacing is $\delta/s \approx 4$. {The flow is incompressible and the density is always scaled with the fluid density, implying that $\rho = 1$.} The simulations are run at a constant flow rate, with the viscosity, $\nu$, adjusted to obtain a friction Reynolds number $Re_\tau = u_\tau \delta/ \nu \approx 185$ {for most of the cases}, where $u_\tau$ is the friction velocity calculated at the canopy tips. In order to ascertain the effects of the Reynolds number on the flow, a simulation at $\Rey_\tau \approx 405$ was also conducted. The simulation parameters are given in table~\ref{tab:DNS_param} for reference. {Scaling with $u_\tau$ and $\nu$ is referred to as in friction or wall units, and scaling with the channel bulk velocity, $U_b$, and $\delta$ is referred to as in outer units}.	
\begin{table}
\begin{center}

%\lineup
\begin{tabular}{clcccccccc}
%\br
           &Case               &$N_x \times N_z$  &$n_x \times n_z$     &{$u_\tau$}  &$\Rey_\tau$ &{$\lambda_f$} & $h^+$     & $s^+$    &{$w^+$}   \\
\hline
%\mr
\multirow{1}{*}{Smooth}  &SC                 &--                         &--                   &{0.064}        &186.3   &--         &--                       & --            &--       \\
\hline
              &S10             &108$\times$54   &12$\times$12        &  {0.071}       &176.7      &{4.6}     &95.5             &10.3       & {5.2}          \\
 \multirow{1}{*}{Fixed}      &S16(H96)      &72$\times$36      &24$\times$24       &{0.088}     &187.7   &{3.1}          &101.8   &16.4  &{8.2}   \\
  \multirow{1}{*}{height} &S24          &48$\times$24    &24$\times$24         & {0.102}       &187.4     &{2.0}       & 101.2    &24.5   &{12.3}  \\
   \multirow{1}{*}{($h^+ \approx 100$)}           &S32          &36$\times$18    &24$\times$24       &{0.112}      &186.4    &{1.5}      &100.7            &32.5        & {16.3}      \\
              &S48            &24$\times$12    &24$\times$24         &   {0.124}      &180.1    &{1.0}          &97.0              &47.2       &  {23.6}      \\
\hline
              &H16            &72$\times$36    &24$\times$24        & {0.071}     &184.7     &{0.5}        &17.2                & 16.1             &{8.1}         \\
\multirow{1}{*}{Fixed}              &H32             &72$\times$36    &24$\times$24        &{0.080}    &188.8    &{1.0}         &34.6               &16.5             &{8.3}          \\
\multirow{1}{*}{spacing}   &H64            &72$\times$36      &24$\times$24      &  {0.086}    &186.2   &{2.0}    &68.1       &16.2     &{8.1}       \\
 \multirow{1}{*}{($s^+ \approx 16$)}   &H96(S16)            &72$\times$36      &24$\times$24      &{0.088}     &185.7  &{3.1}       &101.8      &16.4         & {8.2}             \\
            &H128           &72$\times$36      &24$\times$24       &   {0.086}     &184.8   &{4.1}         &133.2               &16.4          & {8.2}            \\
            
\hline
            &G10    &432$\times$216   &9$\times$9          &   {0.064}     &175.9     &        &10.1         &2.6    & {0.6}         \\
\multirow{1}{*}{Self-similar}      &G20       &216$\times$108    &9$\times$9          &  {0.072}      &190.7   &          &22.2        &5.6     &{1.2}   \\           
\multirow{1}{*}{geometry}     &G40          &108$\times$54     &18$\times$18       &  {0.106}     &188.2     &{0.85}       &43.4        &11.0     & {2.5}    \\           
\multirow{1}{*}{($h/s \approx 4$)}    &G60          &72$\times$36       &18$\times$18       &   {0.127}      &183.3    &       &64.7       &16.0     & {3.6}\\           
                                    &G100           &48$\times$24     &18$\times$18        &  {0.147}      &185.7    &        &97.9      &24.3        &{5.4} \\           
\hline
 \multirow{2}{*}{Different $\Rey_\tau$}   &H32$_{180}$         &72$\times$36     &12$\times$12        & {0.075}    &184.6    &\multirow{2}{*}{{1.0}}         &33.9               &16.1             &{8.1}     \\
                                                 &H32$_{400}$         &162$\times$81     &12$\times$12       &  {0.066}     &399.9  &    &32.0    &15.5    &{7.8}       \\
\hline
%\br
\end{tabular}
\caption{\label{tab:DNS_param} {Simulation parameters. $N_x$ and $N_z$ are the number of rows of canopy elements in the streamwise and spanwise directions, respectively. The number of points used to resolve each period of the canopy in the streamwise and spanwise directions are $n_x$ and $n_z$, respectively. $u_\tau$ is the friction velocity based on the shear at the canopy tips scaled with the channel bulk velocity. $\Rey_\tau$ is the friction Reynolds number based on $u_\tau$ and $\delta$ . The canopy frontal area density, height, spacing and width are $\lambda_f$, $h$, $s$ and $w$, respectively. }}
\end{center}
\end{table}  

The numerical method used to solve the three-dimensional Navier--Stokes equations is adapted from \citet{Fairhall2018}. A Fourier spectral discretisation is used in the streamwise and spanwise directions. The wall-normal direction is discretised using a second-order centred difference scheme on a staggered grid. The grid in the wall-normal direction is stretched to give a resolution $\Delta y^+_{min} \approx 0.33$ at the canopy-tip plane, stretching to $\Delta y^+_{max} \approx 3.3$ at the channel centre. The grid within the canopies preserves the resolution of $\Delta y^+_{min} \approx 0.33$ near the canopy-tip plane, and for the tallest canopies considered stretches to $\Delta y^+_{max} \approx 4$ at the base of the canopy, where the flow is quiescent. {The wall-normal grid distribution for a representative canopy simulation is provided in figure~\ref{fig:grid} in appendix~\ref{appA} for reference.} To resolve the element-induced flow while avoiding excessive computational costs, the domain is divided into three blocks in the wall-parallel directions \citep{Garcia-Mayoral2011,Fairhall2018,Abderrahaman2019}. In the central block, the resolutions in the streamwise and spanwise directions are $\Delta x^+ \approx 6$ and $\Delta z^+ \approx 3$, respectively, sufficient to resolve the turbulent eddies. The blocks including the canopy elements and the roughness sublayer have a finer resolution than the central block. In the fine blocks, the limiting resolution is not the one required to resolve the turbulent scales, but that required to resolve the obstacles or the element-induced flow. The resolutions in these blocks are given in table~\ref{tab:DNS_param}. The height of the fine blocks is chosen such that the element-induced flow decays to zero well within the fine-block region, and this is verified a posteriori. The time advancement is carried out using a three-step Runge--Kutta method with a fractional step, pressure correction method to enforce continuity \citep{Le1991}

\begin{eqnarray}
\label{eq:disc_NS}\left[\mathrm{I} - \Delta t \frac{\beta_k}{\Rey} \mathrm{L}\right] \mathbf{u}^n_k & = & \mathbf{u}^n_{k-1} +  \Delta t \left[ \frac{\alpha_{k}}{\Rey}\mathrm{L} \mathbf{u}^n_{k-1} - \gamma_{k}\mathrm{N}(\mathbf{u}^n_{k-1}) - \right.\nonumber\\
&&\left. \zeta_k \mathrm{N}(\mathbf{u}^n_{k-2}) - (\alpha_k + \beta_k)\mathrm{G}(p^n_k) \vphantom{ \frac{\alpha_{k}}{\Rey}}\right], k\in[1,3],\\ 
\mathrm{DG}(\phi^n_k) & = &  \frac{1}{(\alpha_k + \beta_k) \Delta t} \mathrm{D}(\mathbf{u}^n_k), \\
\mathbf{u}^n_{k+1} & = &  \mathbf{u}^n_k - (\alpha_k + \beta_k)\Delta t G(\phi^n_k),\\
p^n_{k+1} & = &  p^n_k + \phi^n_k,
\end{eqnarray}
where $\mathrm{I}$ is the identity matrix and L, D and G are the Laplacian, divergence and gradient operators respectively. N is the advective term which is dealiased using the $2/3$-rule \citep{Canuto2012}. The Runge-Kutta coefficients, $\alpha_k$, $\beta_k$, $\gamma_k$ and $\zeta_k$, for each substep, $k$, are taken from from \citet{Le1991}. The time step is $\Delta t$.

\begin{figure}
        \centering
		\vspace{5mm} \includegraphics[scale=0.59,trim={0mm 0mm 0mm 0mm},clip]{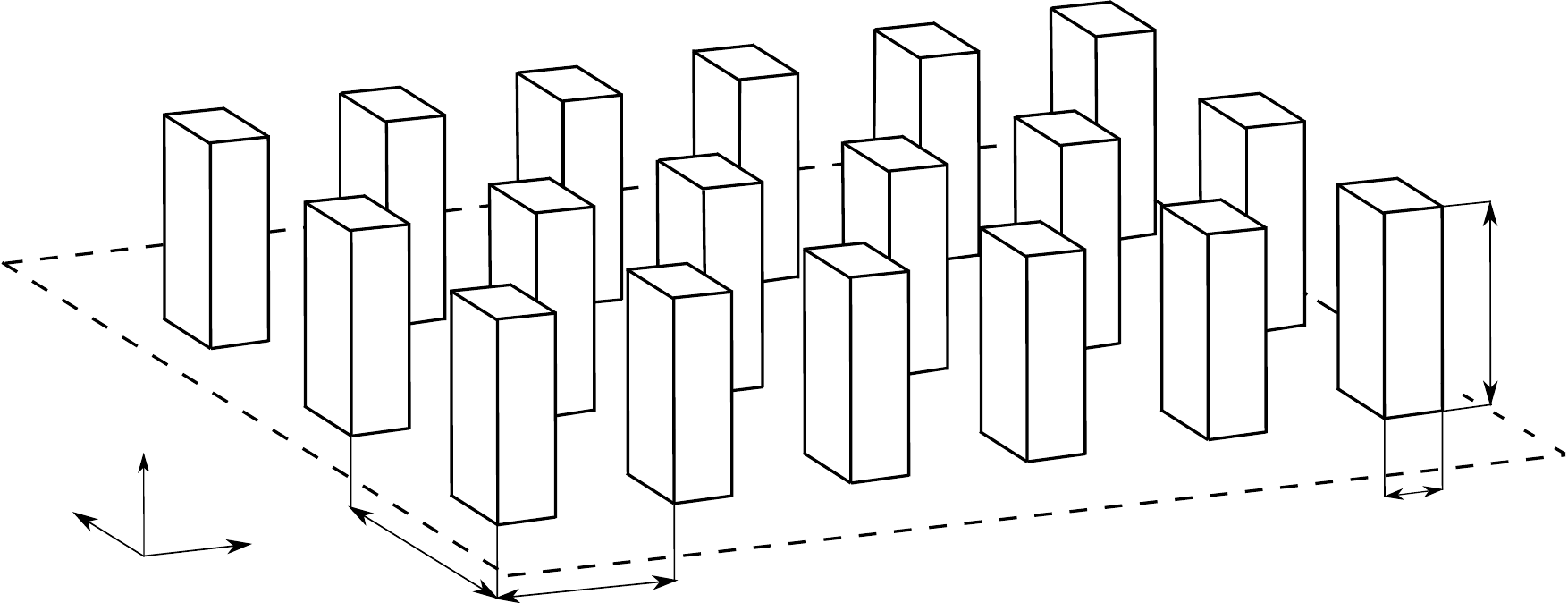}%
  		\mylab{-0.4cm}{1.95cm}{$h$}%
  		\mylab{-1.12cm}{0.45cm}{$w$}%
  		\mylab{-6.6cm}{-0.15cm}{$s$}%
  		\mylab{-7.9cm}{0.15cm}{$s$}%
  		\mylab{-9cm}{0.1cm}{$x$}%
  		\mylab{-9.8cm}{0.9cm}{$y$}%
  		\mylab{-10.2cm}{0.3cm}{$z$}%
  		\caption{Schematic of the canopy layouts considered in the present study. The canopies are characterised by their element height, $h$, the element width, $w$, and the element spacing, $s$. Note that the element have a square top-view cross section.}%
   	     \label{fig:canopy_schem}
\end{figure}

The canopy elements are represented using an immersed-boundary method adapted from \citet{Garcia-Mayoral2011}. {Further details about the immersed boundary method and validation studies are provided in appendix~\ref{appA}.} The parameters of the different simulations conducted are summarised in table~\ref{tab:DNS_param}. The simulation denoted by `SC' is of a turbulent channel flow with smooth walls. The canopy-flow simulations are divided into three groups. The canopy elements studied in each group are prismatic, with a square top-view cross section, and their arrangement is illustrated in figure~\ref{fig:canopy_schem}. The first group, denoted by the prefix `S', consists of canopies with a fixed height, $h^+ \approx 96$, and element spacings ranging from $s^+ \approx 10$ to $48$. The second group, marked by the prefix `H', consists of canopies with a fixed element spacing, $s^+ \approx 16$, and element heights ranging from $h^+ \approx 16$ to $128$. The element width-to-spacing ratio for the canopies of S and H is $w/s = 1/2$. The final group, denoted by the prefix `G', consists of self-similar elements with a fixed height-to-spacing ratio $h/s \approx 4$, and $w/s = 2/9$. The heights for the canopies of G range from $h^+ \approx 10$ to $100$, with the element spacings varying in proportion to their height. {These canopies have a constant $\lambda_f = 0.85$ and are used to study the effect of changing the canopy size for a fixed geometry}. Two additional simulations, H32$_{180}$ and H32$_{400}$, are conducted to check the dependence of the results on the friction Reynolds number. The canopy geometries for both these simulations have $s^+ \approx 16$, $h^+ \approx 32$ and $w/s = 1/2$, with friction Reynolds numbers $\Rey_\tau \approx 180$ and $400$. We also conducted several simulations to assess whether the wall-parallel resolutions used in the simulations are sufficient to resolve the element-induced flow. The simulation S24 was run at resolutions of $12$, $24$ and $36$ points per element spacing, and G100 at $9$, $18$ and $27$ points per element spacing. Different resolution sets are used for the geometries of S and G as they have different element width-to-spacing ratios. The rms velocity fluctuations obtained from these simulations are portrayed in appendix~\ref{appA}. The simulation results are grid independent at a resolution of $24$ points per element spacing for the geometry of case S24, and $18$ points per spacing for that of case G100. The simulations with $9$ and $12$ points per spacing tend to under-predict the fluctuations within the canopies, with the maximum deviation observed in the wall-normal fluctuations of $20\%$ within the canopy. This discrepancy reduces to $4\%$ outside the canopy. These resolutions are only used for the densest canopy cases, where the fluctuations within the canopy are already very small, and therefore, higher resolution simulations would not change the trends observed. The higher Reynolds number simulation, case H32$_{400}$, is also simulated using $12$ points per spacing. For this simulation, using a higher resolution would be computationally restrictive. Note that the same resolutions are used for cases H32$_{180}$ and H32$_{400}$ to avoid grid related discrepancies in the comparison of their results.

\subsection{Reynolds number effect}\label{sec:Rey_effects}
 \begin{figure}
	\centering
		\subfloat{%
%  		 \tikzsetnextfilename{urms_Re_var}
  		\hspace{-2.3mm}\includegraphics[scale=1]{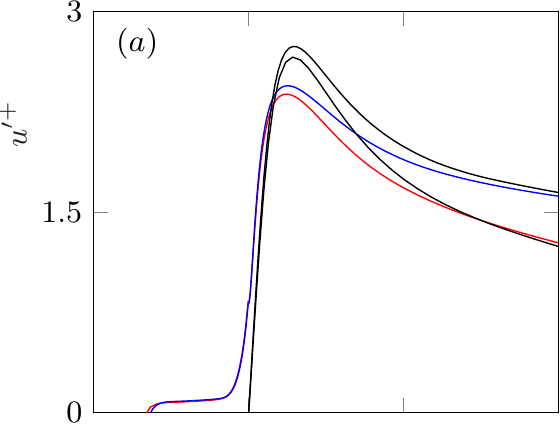}
  		}%
  		\subfloat{%
%  		 \tikzsetnextfilename{vrms_Re_var}
  			\hspace{3mm}\includegraphics[scale=1]{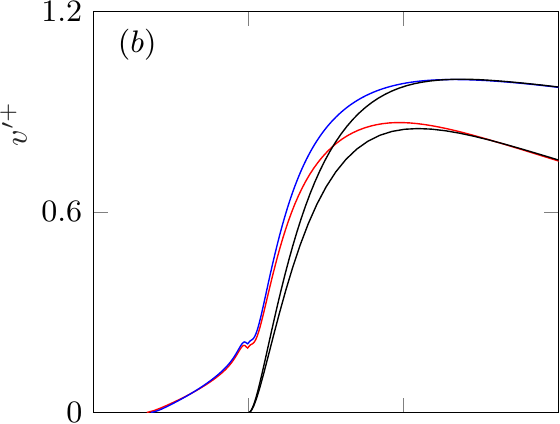}
  		}%
  		
  		\vspace{0mm}\subfloat{%
% 		\tikzsetnextfilename{wrms_Re_var}
 		\hspace{0mm}\includegraphics[scale=1]{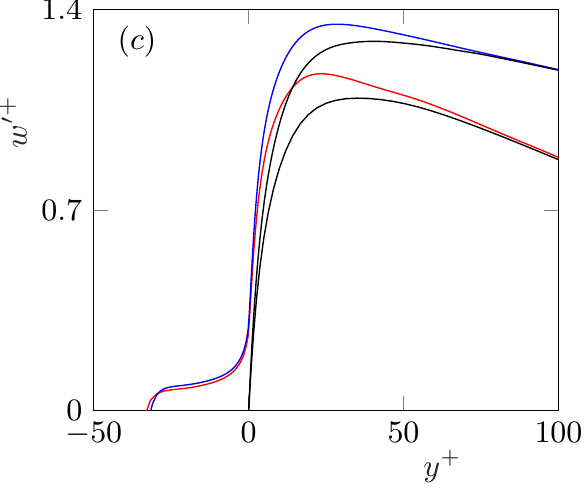}
 		}%
 		\subfloat{%
% 		\tikzsetnextfilename{uvrms_Re_var}
 		\hspace{0mm}\includegraphics[scale=1]{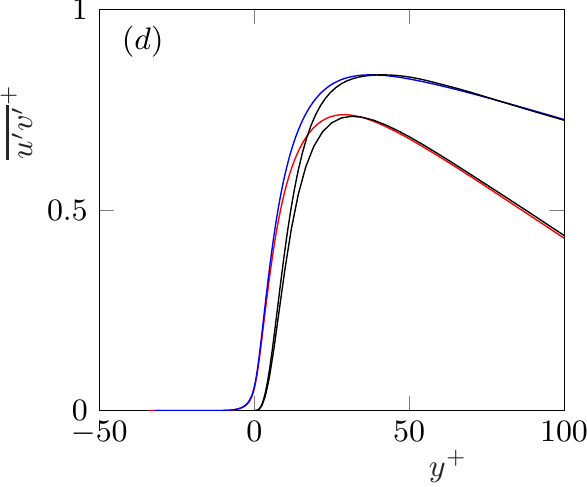}
 		}%
 		 \caption{Rms velocity fluctuations and Reynolds shear stresses for cases H32$_{180}$ in red and H32$_{400}$ in blue. The black lines represent the corresponding smooth-wall cases. The data for the smooth-wall simulations at $\Rey_\tau \approx 400$ is taken from \citet{Moser1999}.}
 		  	\label{fig:stats_Re}	
\end{figure}	

To analyse the influence of the Reynolds number in our subsequent DNSs, we compare the results of cases H32$_{180}$ and H32$_{400}$, which have the same canopy height and spacings in friction units, but different friction Reynolds numbers. The velocity fluctuations and the Reynolds shear stresses within the canopy, and above it up to a height of $y^+\approx 10$, of these simulations essentially collapse, as shown in figure~\ref{fig:stats_Re}. This suggests that the flow in the region near the canopy-tip plane scales in friction units, similar to the near-wall region in smooth-wall flows \citep{Moser1999}. Scaling in friction units over conventional rough surfaces has also been noted by \citet{Chan2015}. Beyond $y^+ \gtrsim 10$, we observe that the magnitude of the peaks in the fluctuations and the Reynolds shear stresses are larger for case H32$_{400}$ compared to case H32$_{180}$. The increase in magnitude of the near-wall peaks in the velocity fluctuations at friction Reynolds numbers larger than $Re_\tau \approx 180$ is consistent with that observed in smooth-wall flows \citep{Moser1999,Sillero2013}, also included in figure~\ref{fig:stats_Re} for reference. Further away from the canopy tips, at $y^+ > 50$, the rms velocity fluctuations from the canopy simulations coincide with those from the smooth-wall simulations at their corresponding Reynolds numbers, which indicates the recovery of outer-layer similarity.
\begin{figure}
\centering
		\vspace{6mm} \includegraphics[scale=1.0,trim={8mm 0mm 0mm 0mm},clip]{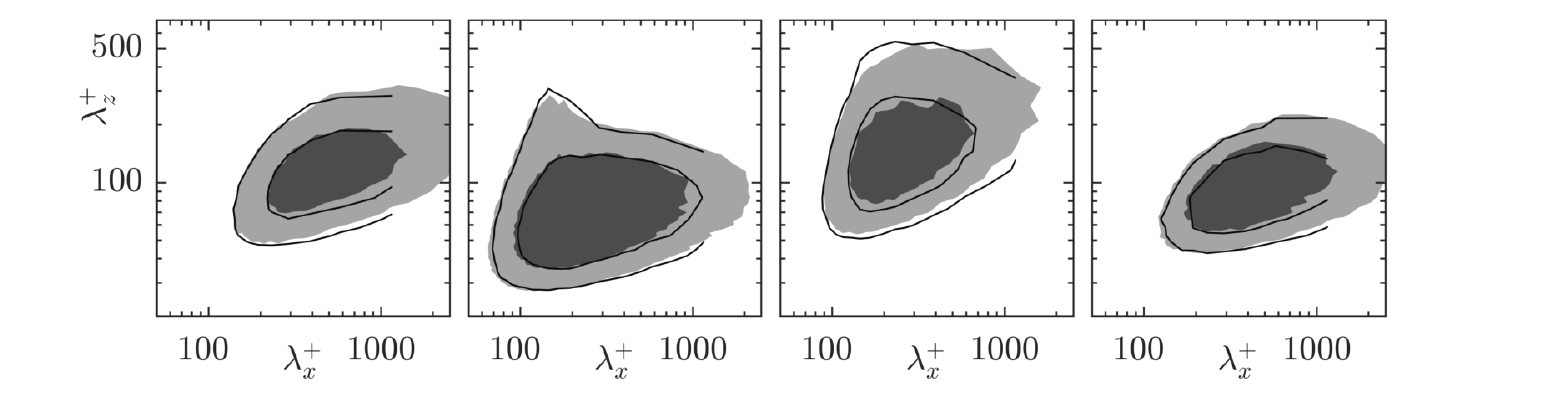}%
         \mylab{-14.2cm}{3.5cm}{(\textit{a})}%
  		\mylab{-11.0cm}{3.5cm}{(\textit{b})}%
  		\mylab{-7.8cm}{3.5cm}{(\textit{c})}%
  		\mylab{-4.7cm}{3.5cm}{(\textit{d})}%
  		\mylab{-13.3cm}{4.0cm}{$k_x k_z E_{uu}$}%
   		 \mylab{-10.2cm}{4.0cm}{$k_x k_z E_{vv}$}%
   		 \mylab{-7.0cm}{4.0cm}{$k_x k_z E_{ww}$}% 	
   		 \mylab{-3.8cm}{4.0cm}{$k_x k_z E_{uv}$}% 	
  		\caption{Pre-multiplied spectral energy densities for cases H32$_{180}$ (line contours) and H32$_{400}$ (shaded contours), normalised by the respective rms values, at a height $y^+ \approx 15$. Contours from ($a$--$d$) are in increments of $0.075$, $0.06$, $0.07$ and $0.1$, respectively.}%
   	     \label{fig:spectra_h32_Re_var}
\end{figure} 
In addition to the rms fluctuations being similar for these simulations, the distribution of energy in different scales is also similar. This is illustrated by the pre-multiplied spectral energy densities at $y^+ \approx 15$, portrayed in figure~\ref{fig:spectra_h32_Re_var}. This height roughly corresponds to the location where the magnitude of the fluctuations peaks in smooth-wall flows \citep{Jimenez1999}. The results of H32$_{180}$ and H32$_{400}$ suggest that the effect of the canopy scales in friction units, and therefore the results presented in the following sections for flows at $\Rey_\tau \approx 180$ should also be relevant for higher Reynolds number flows.

\begin{figure}
	\centering
  		\vspace{2mm}\subfloat{%
  	    \includegraphics[scale=1,trim={20mm 39mm 18mm 22mm},clip]{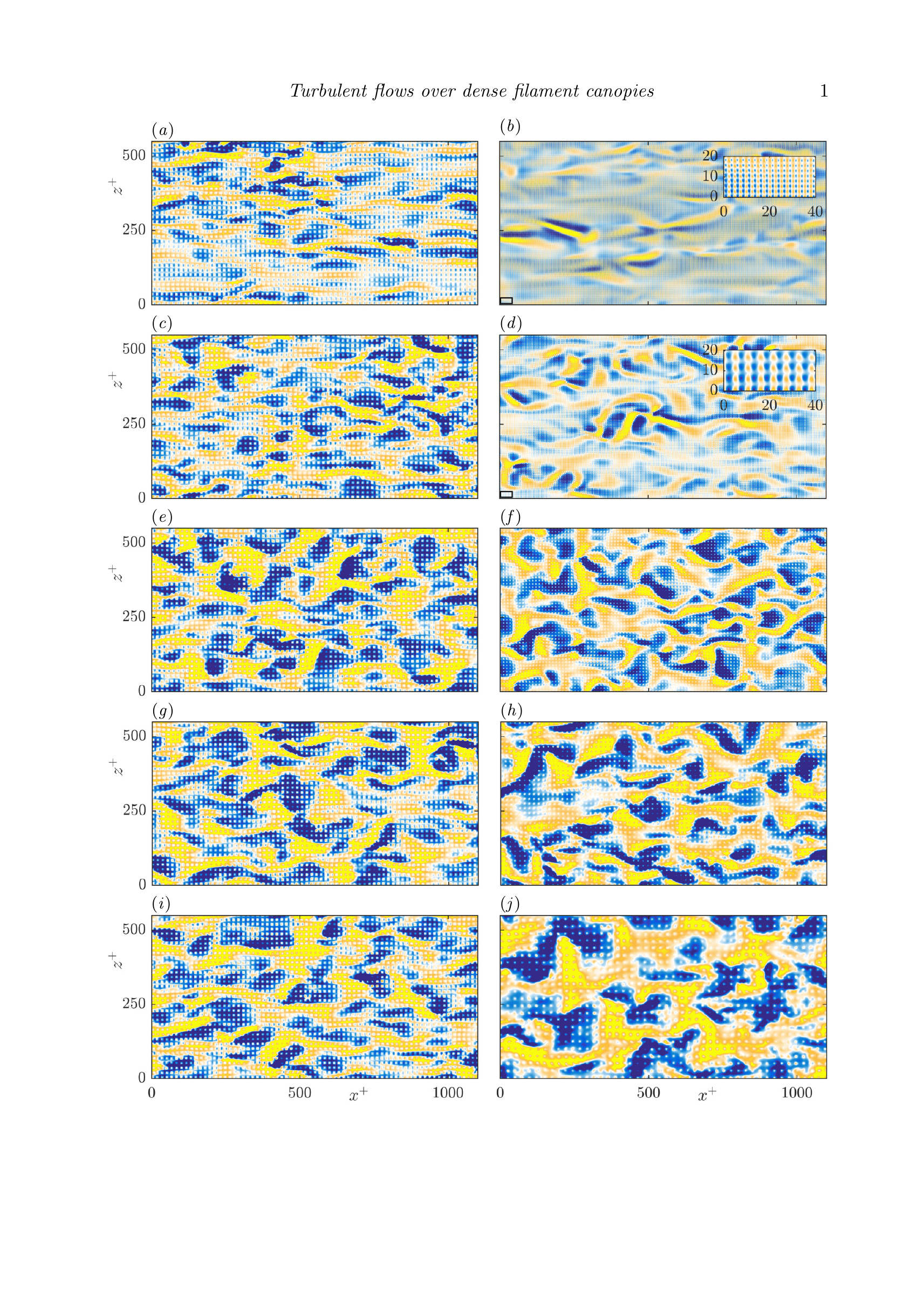}
  		}%
 		\caption{Instantaneous realisations of the wall normal velocity at $y \approx 0.1 s$, normalised by $u_\tau$. From top to bottom, the left column represents cases H16 to H128; and the right column, cases G10 to G100. The insets in ($b$) and ($d$) provide a magnified view of the region in the bottom left corner of these panels, marked with a black rectangle. The clearest and darkest colours represent intensity $\pm 0.4$ in the left column and, from top to bottom, $\pm(0.2, 0.4, 0.8, 0.8, 1.0)$ in the right column.}%
   	     \label{fig:v_snap_G}
\end{figure}

   \begin{figure}
	\centering
		\hspace{0.9mm}\subfloat{%
%  		 \tikzsetnextfilename{ucrms_S}
  		\includegraphics[scale=1]{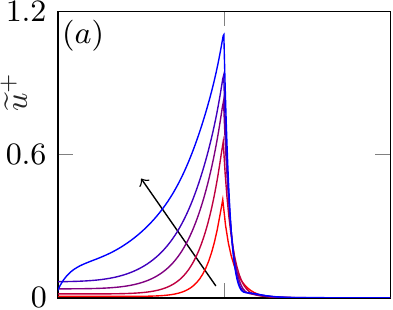}
  		}%
  		\hspace{2.7mm}\subfloat{%
%  		 \tikzsetnextfilename{ucrms_H}
  		\includegraphics[scale=1]{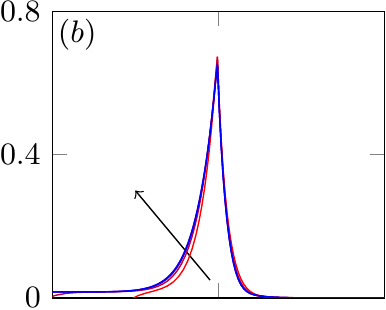}
  		}%
  		\hspace{1.2mm} \subfloat{%
%  		 \tikzsetnextfilename{ucrms_G}
  		\includegraphics[scale=1]{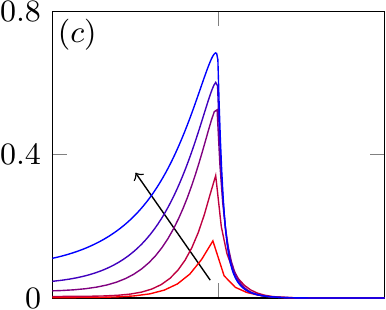}
  		}%
  		
  		\vspace{0mm}\subfloat{%
%  		 \tikzsetnextfilename{vcrms_S}
  		\includegraphics[scale=1]{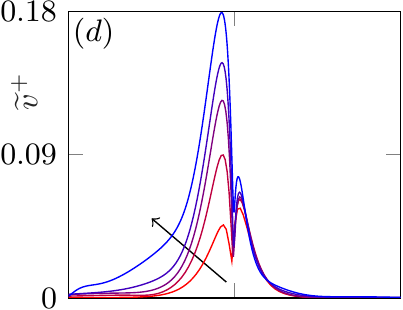}
  		}%
  	\hspace{0mm}	\subfloat{%
%  		 \tikzsetnextfilename{vcrms_H}
  		\includegraphics[scale=1]{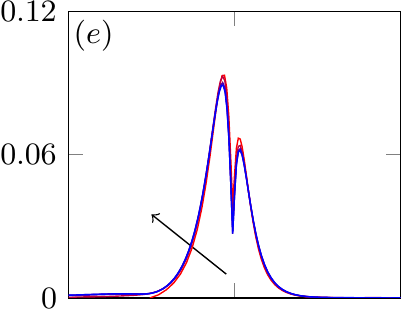}
  		}%
  		\hspace{0.5mm}\subfloat{%
%  		 \tikzsetnextfilename{vcrms_G}
  		\includegraphics[scale=1]{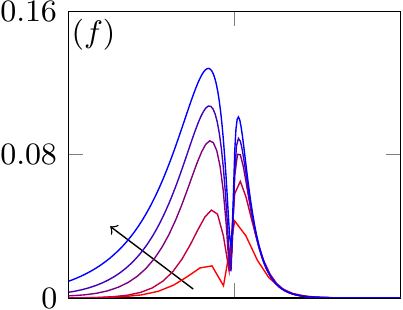}
  		}%
  		
   	\vspace{3mm} \hspace{0.4mm}\subfloat{%
% 	 \tikzsetnextfilename{wcrms_S}
  		\includegraphics[scale=1]{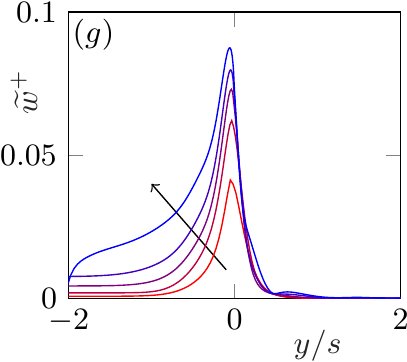}
 		}%
  \hspace{-0.3mm}	\subfloat{%
% 	 \tikzsetnextfilename{wcrms_H}
  		\includegraphics[scale=1]{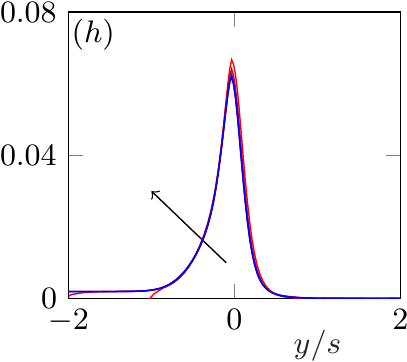}
 		}%
   \hspace{-0.9mm}	\subfloat{%
% 	 \tikzsetnextfilename{wcrms_G}
  		\includegraphics[scale=1]{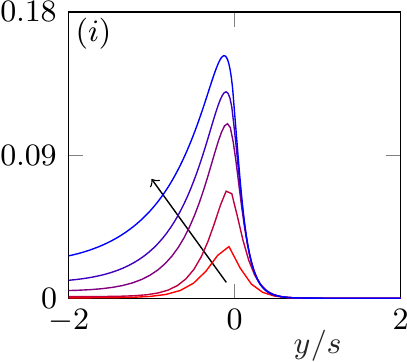}
 		}%

 	 \caption{Root-mean-square velocity fluctuations of the element-induced flow. The lines from red to blue, {indicated by the direction of the arrows}, represent ($a$,$d$,$g$) cases S10 to S48; ($b$,$e$,$h$) cases H16 to H128; and ($c$,$f$,$i$) cases G10 to G100.}
 		  	\label{fig:stats_cond}	
\end{figure}

\section{Effect of canopy parameters on the surrounding turbulence}\label{sec:turb_fluct}

{In this section, we discuss the results obtained from the DNSs, aiming to characterise the effect of the canopy parameters on both the Kelvin--Helmholtz-like instability and the fluctuating flow within and above the canopies.}

\subsection{{Height of the roughness sublayer}}\label{subsec:RSL}
{Before discussing the effect of the canopy on the overlying flow, we first define the extent of the region affected by the canopy, that is, the height of the roughness sublayer.} Over conventional rough surfaces, with heights comparable to or smaller than their spacings, the height of the roughness sublayer is generally observed to be a function of the roughness height \citep{Raupach1991,Flack2007,Abderrahaman2019}. \citet{Jimenez2004} reviewed the effect of various roughness geometries on turbulent flows and noted that, in flows over closely packed spanwise aligned grooves, the flow within each groove would be isolated from the overlying flow due to the `sheltering' effect of the preceding obstacle. The overlying flow in this case would not interact with the full height of the groove. This sheltering effect was also noted by \citet{Sadique2017} for high-aspect-ratio prismatic roughness and by \citet{Macdonald2018} for spanwise aligned grooves with large spacings, and has been used to model cuboidal roughness by \citet{Yang2016}. As the element spacings of the canopies studied here are small, this sheltering effect should result in the overlying flow only interacting with the region near the canopy-tip plane. In order to determine the height of this region, we examine the element-coherent flow induced by the canopy elements. The footprint of the element-induced flow can be observed in the instantaneous realisations of the velocity above the canopy-tip plane, portrayed for the canopies of families H and G in figure~\ref{fig:v_snap_G}. We isolate the element-induced flow using the standard triple decomposition of \citet{Reynolds1972}
\begin{eqnarray}
\boldsymbol{u} = \boldsymbol{U}  + \boldsymbol{u'},\\
\boldsymbol{u'}= \boldsymbol{\widetilde{u}} + \boldsymbol{u''},
\end{eqnarray}
where $\boldsymbol{u}$ is the full velocity, $\boldsymbol{U}$ is the mean velocity obtained by averaging the flow in time and space, and $\boldsymbol{u'}$ is the full space- and time-fluctuating signal. The latter is decomposed into the element-induced, dispersive velocity, $\boldsymbol{\widetilde{u}}$, which is obtained by {ensemble-averaging} the flow in time alone, and the {element-}incoherent fluctuating velocity $\boldsymbol{u''}$, {which includes the contributions from the background-turbulence and the Kelvin--Helmholtz-like instability. The rms fluctuations of $\boldsymbol{\widetilde{u}}$, therefore, result from fluctuations in space alone.} 

We observe that the element-induced fluctuations, for all the canopies studied here, decay exponentially above the canopy-tip plane, and become negligible at a height of one element spacing above regardless of the canopy depth, as shown in figure~\ref{fig:stats_cond}. {This suggests that the height of influence of the element-induced flow is determined by the spacing between the elements rather than their height.}

\begin{figure}
	\centering
		\subfloat{%
%  		 \hspace{0mm}\tikzsetnextfilename{urms_S}
  		\hspace{-2mm}\includegraphics[scale=1, trim = 0cm 0cm 0cm 0cm,clip]{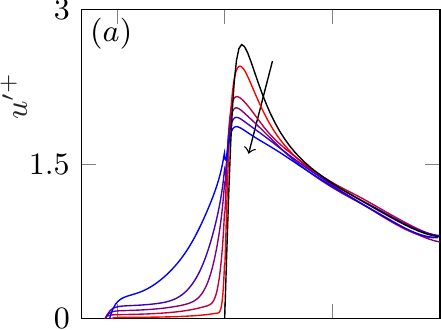}
  		}%
  		\hspace{3.5mm}\subfloat{%
%  		 \tikzsetnextfilename{urms_H}
  		\includegraphics[scale=1, trim = 0cm 0cm 0cm -0.07cm,clip]{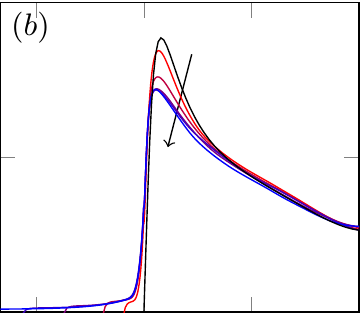}
  		}%
  		\hspace{2.2mm} \subfloat{%
%  		 \tikzsetnextfilename{urms_G}
  		\includegraphics[scale=1, trim = 0cm 0cm 0cm -0.07cm,clip]{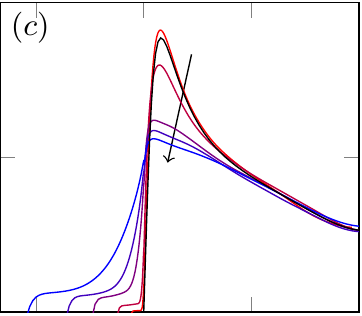}
  		}%
  		
  		\vspace{-1.5mm}\subfloat{%
%  		\hspace{0mm} \tikzsetnextfilename{vrms_S}
  		\hspace{-2mm}\includegraphics[scale=1, trim = 0cm 0cm 0cm 0cm,clip]{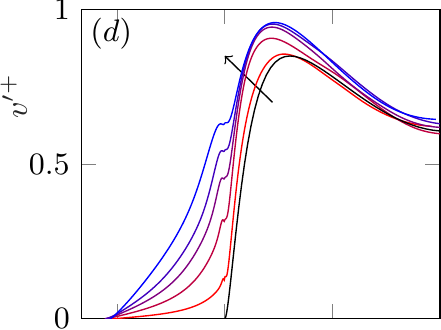}
  		}%
           \subfloat{%
%  		 \tikzsetnextfilename{vrms_H}
  		\hspace{3.5mm}\includegraphics[scale=1, trim = 0cm 0cm 0cm -0.07cm,clip]{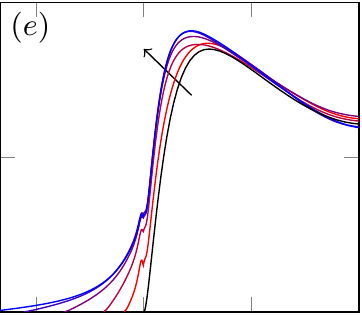}
  		}%
             \subfloat{%
%  		 \tikzsetnextfilename{vrms_G}
  		\hspace{3.3mm}\includegraphics[scale=1, trim = 0cm 0cm 0cm -0.07cm,clip]{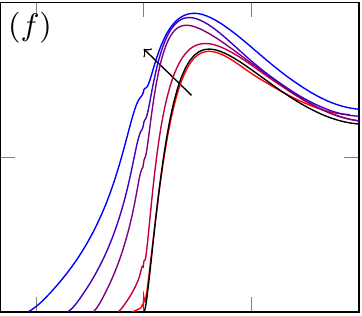}
  		}%
  		
   	\vspace{-1.5mm}\subfloat{%
% 	 \hspace{0mm}\tikzsetnextfilename{wrms_S}
  		\includegraphics[scale=1, trim = 0cm 0cm 0cm 0cm,clip]{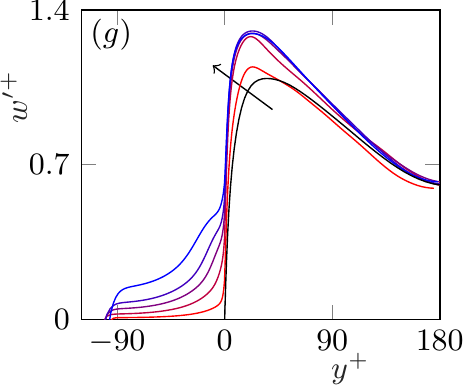}
 		}%
  \hspace{0mm}	\subfloat{%
% 	 \tikzsetnextfilename{wrms_H}
  		\includegraphics[scale=1, trim = 0cm 0cm 0cm -0.07cm,clip]{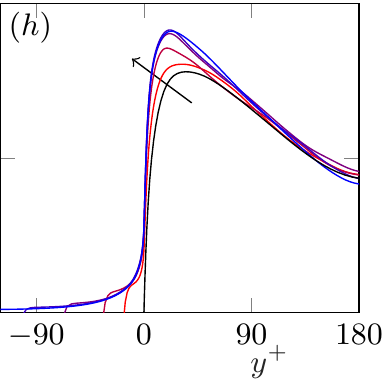}
 		}%
   \hspace{0mm}	\subfloat{%
% 	 \tikzsetnextfilename{wrms_G}
  		\includegraphics[scale=1, trim = 0cm 0cm 0cm -0.07cm,clip]{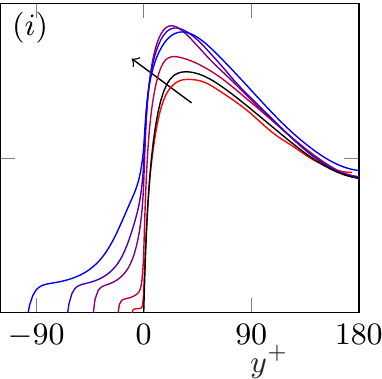}
 		}%
 			 \caption{Rms velocity fluctuations within and above the canopies. The lines from red to blue, {indicated by the direction of the arrows},  represent ($a$,$d$,$g$) cases S10 to S48; ($b$,$e$,$h$) cases H16 to H128; and ($c$,$f$,$i$) cases G10 to G100. The black lines represent the smooth-wall case, SC.}
 		  	\label{fig:stats_all}	
\end{figure}

\begin{figure}
\centering
		 \vspace*{0.3cm}\includegraphics[scale=1.0,trim={7mm 0mm 0mm 0mm},clip]{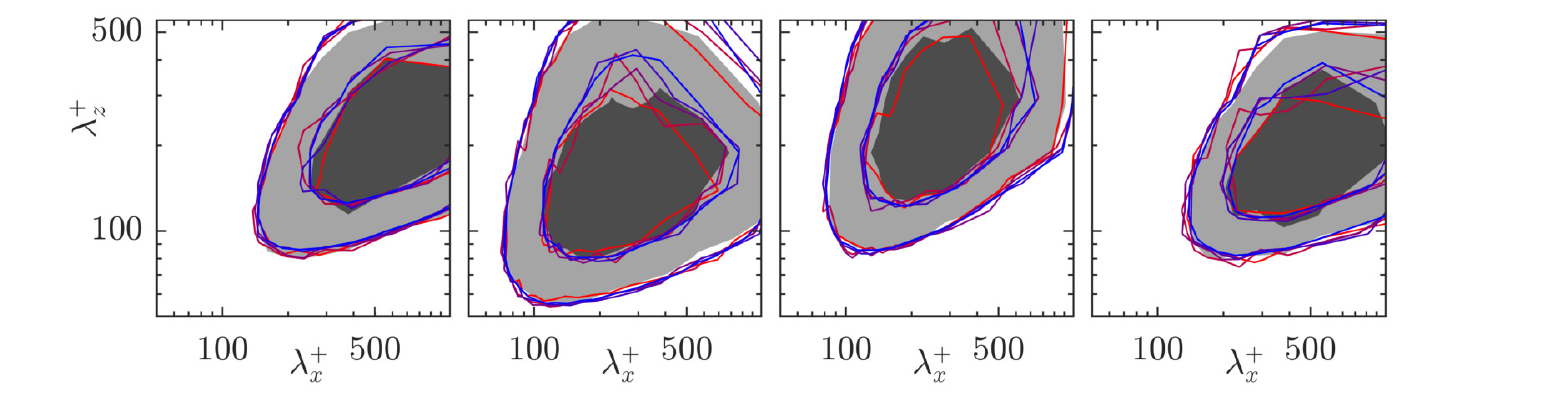}%
		\mylab{-14.2cm}{3.5cm}{(\textit{a})}%
  		\mylab{-11.0cm}{3.5cm}{(\textit{b})}%
  		\mylab{-7.8cm}{3.5cm}{(\textit{c})}%
  		\mylab{-4.7cm}{3.5cm}{(\textit{d})}%
  		\mylab{-13.3cm}{4.0cm}{$k_x k_z E_{uu}$}%
   		 \mylab{-10.2cm}{4.0cm}{$k_x k_z E_{vv}$}%
   		 \mylab{-7.0cm}{4.0cm}{$k_x k_z E_{ww}$}% 	
   		 \mylab{-3.8cm}{4.0cm}{$k_x k_z E_{uv}$}% 	
  		\caption{Pre-multiplied spectral energy densities at $y^+ \approx 90$, with line contours from red to blue representing cases S10 to S48, normalised by their respective $u_\tau$. The filled contours represent the smooth-wall case, SC. The contours in ($a$--$d$) are in increments of $0.11$, $0.04$, $0.06$ and $0.04$, respectively.}%
   	     \label{fig:spectra_h90_OLS}
\end{figure}

Even though the element-induced fluctuations only extend to one element spacing above the canopy-tip plane, their influence on the background-turbulence extends to a height of approximately $2$--$3$ element spacings, {as can be observed in figure~\ref{fig:stats_all}}. \citet{Abderrahaman2019} observed a similar effect over conventional cubical roughness, where the element-induced fluctuations only extended to $y \approx h$, but the effect of the roughness on the overlying flow extended to $y \approx 3h$ above them. At heights of $y/s > 2$--$3$ above the canopy-tip plane, the full rms velocity fluctuations collapse with those of smooth-wall turbulence, as shown in figure~\ref{fig:stats_all}, which is indicative of the recovery of outer-layer similarity. This is verified by a comparison of the pre-multiplied spectral energy densities of the canopy and smooth-wall cases in figure~\ref{fig:spectra_h90_OLS}, which shows that the energy densities of the canopies of family S collapse with those of the smooth-wall case for $y^+ \gtrsim 90$. This corresponds to a height of about $2s$ for case S48, the canopy with the largest spacing. Although not shown, the pre-multiplied spectral energy densities of the canopies of families H and G collapse with the smooth-wall spectra for $y/s \gtrsim 3$ as well. {Previous canopy studies have proposed that the influence of the canopy elements on the flow above them is set by the wall-normal extent of the Kelvin--Helmholtz-like instability, which is determined by the canopy shear-layer thickness \citep{Ghisalberti2009,GhisalbertiNepf2009,Nepf2012}. It will be demonstrated in \S\ref{sec:stability_analysis} that the shear-layer thickness of the present canopies also depends on the element spacing.}

\begin{figure}
	\centering
		\subfloat{%
%  		 \tikzsetnextfilename{U_S}
  		\hspace{-1.2mm}\includegraphics[scale=1, trim=0mm 0mm 0mm 0mm,clip]{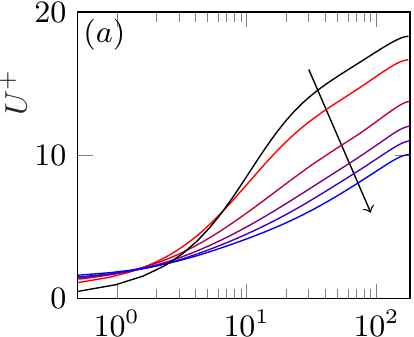}
  		}%
  		\subfloat{%
%  		 \tikzsetnextfilename{U_H}
  		\hspace{4.5mm}\includegraphics[scale=1, trim=0mm 0mm 0mm -1mm,clip]{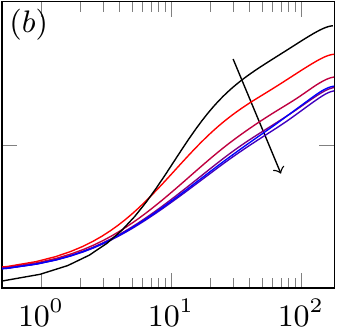}
  		}%
  		\subfloat{%
%  		 \tikzsetnextfilename{U_G}
  		\hspace{3.5mm} \includegraphics[scale=1, trim=0mm 0mm 0mm -1mm,clip]{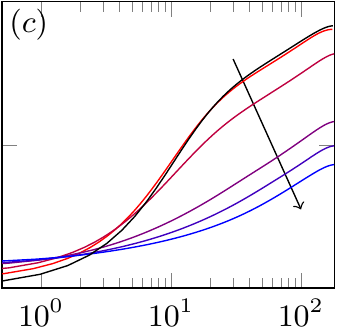}
  		}%
  		
  		\vspace{0mm}\subfloat{%
%  		 \tikzsetnextfilename{uv_S}
  		\includegraphics[scale=1, trim=0mm 0mm 0mm 0mm,clip]{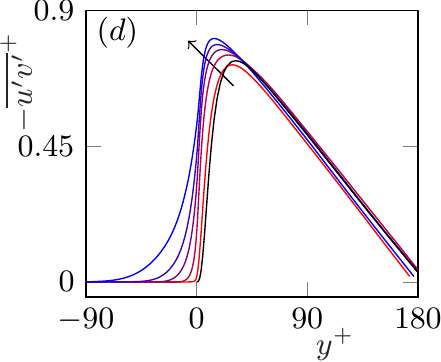}
  		}%
  	\subfloat{%
%  		 \tikzsetnextfilename{uv_H}
  		\hspace{0mm}\includegraphics[scale=1, trim=0mm 0mm 0mm -0.7mm,clip]{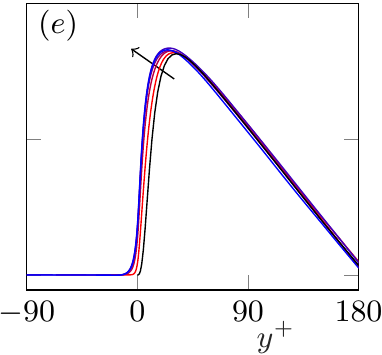}
  		}%
  		\subfloat{%
%  		 \tikzsetnextfilename{uv_G}
  		\hspace{0mm}\includegraphics[scale=1, trim=0mm 0mm 0mm -0.7mm,clip]{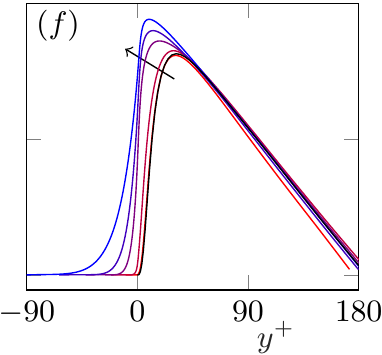}
  		}%
 			 \caption{Profiles  of the ($a$--$c$) streamwise mean velocity and ($d$--$f$) Reynolds shear stresses. The lines from red to blue, {indicated by the direction of the arrows},  represent ($a$,$d$) cases S10 to S48; ($b$,$e$) cases H16 to H128; and ($c$,$f$) cases G10 to G100. The black lines represent the smooth-wall case, SC.}
 		  	\label{fig:U_uv}	
\end{figure}

\subsection{{Effect of element height and spacing}}\label{subsec:fluc}
{In this section, we discuss the effect of the element height and spacing on the element-induced and on the full velocity fluctuations, both within and above the canopy. We observe that the element-induced fluctuations are largest near the canopy-tip plane and decay below it, as shown in figure~\ref{fig:stats_cond}, because they are obstructed by the canopy elements. This effect is more intense for smaller element spacings, and eventually results in the fluctuations vanishing completely well above the canopy floor for the simulations with the smallest spacings in families S and G. For the canopies of family H, which have a constant element spacing, the change in element height does not have a noticeable effect on the element-induced fluctuations, as shown in figures~\ref{fig:stats_cond}($b,e,h$). For the canopies of families S and G, the intensity of the element-induced velocity fluctuations within the canopy increases with element spacing, when scaled using either the friction velocity or the channel bulk velocity. These results suggest that, for canopy elements with a given width, the magnitude of the element-induced fluctuations is governed mainly by the element spacing.}

{As discussed in \S\ref{subsec:RSL},} in canopies with very small element spacing the height of the roughness sublayer is small, and we would expect such canopies not to disrupt the overlying turbulence significantly, regardless of their depth. In the literature on conventional roughness, small roughness elements that have a negligible effect on the overlying turbulent flow are termed `hydraulically-smooth', as the flow over them remains essentially smooth-wall like \citep{Nikuradse1933,Raupach1991}. Roughness elements with a characteristic size of a few wall-units, $h^+ \lesssim 5$, typically fall into the hydraulically smooth category \citep{Raupach1991,Jimenez2004,Flack2007}. Of the canopies studied here, we observe that the overlying flow for canopy G10, which has an element spacing of $s^+ \approx 2.6$, is essentially smooth-wall-like above the canopy-tip plane. This is evidenced by the collapse of the rms velocity fluctuations, Reynolds shear stresses, and the mean velocity profile of this case with those of the smooth-wall case, as shown in figures~\ref{fig:stats_all}($c,f,i$)~and~\ref{fig:U_uv}($c,f$). In addition, the magnitude of the velocity fluctuations below the canopy-tip plane is negligible. This suggests that the overlying turbulent flow essentially perceives the canopy-tip plane as an impermeable wall, and has little or no interaction with the canopy region below this plane. 

For canopies with larger element spacings, we begin to observe deviations from smooth-wall-like behaviour in the overlying flow. Above the canopy-tip plane, an increase in the element spacing causes a  reduction in the intensity of the streamwise velocity fluctuations and an increase in the intensity of the wall-normal and spanwise ones, as can be observed in figure~\ref{fig:stats_all} for the canopies of families S and G. For canopies with large element spacings, such as those of S48 and G100, the peak in $u'$ typical of smooth-wall flows, is significantly reduced. {These changes in the velocity fluctuations are accompanied by a reduction in the streamwise coherence in the flow with increasing element spacing as can be observed in the instantaneous realisations of the wall-normal velocity for the canopies of family G, portrayed in figure~\ref{fig:v_snap_G}. Near-wall turbulence over smooth walls is characterised by streaks and quasi-streamwise vortices, which are predominantly streamwise-coherent \citep{Kline1967,Jimenez1999}. The decrease in $u'$ and increase in $v'$, $w'$ above the canopy with increasing element size is also commonly reported over conventional rough surfaces \citep{Ligrani1986,Orlandi2006}. Several authors have attributed these changes in the velocity fluctuations and the loss of streamwise coherence to the roughness elements modifying the near-wall cycle and turbulence becoming more `isotropic' \citep{Jimenez2004,Flores2006,Flack2007,Abderrahaman2019}. We also observe an increase in the Reynolds shear stresses above the canopy tip plane with increasing element spacing, shown in figures~\ref{fig:U_uv}($d,f$), with an associated increase in the drag. The drag increase caused by rough surfaces is generally expressed in terms of the downward shift in the logarithmic region of the mean-velocity profile compared to that for a smooth wall \citep{Hama1954}. This shift can be observed for the canopies of families S and G in figures~\ref{fig:U_uv}($a,c$).}

{Focusing now on the flow within the canopy, increasing the element spacing results in an increase in the magnitude of all the components of the full velocity fluctuations, as shown in figure~\ref{fig:stats_all}, which is consistent with the observations of \citet{Green1995}, \citet{Novak2000}, \citet{Poggi2004} and \citet{Pietri2009}}. The wall-parallel velocity fluctuations, $u'$ and $w'$, decay rapidly below the canopy-tip plane, and their magnitude reaches a plateau in the core of the canopy, before dropping again near the canopy base to meet the no-slip condition. {The abrupt changes in the velocity fluctuations near the element tips are typical of textures with perfectly flat and aligned tips and have also been observed over conventional cuboidal rough surfaces \citep{Leonardi2010, Abderrahaman2019} and permeable substrates \citep{Kuwata2017}. However, this effect would likely be smeared out over canopies with irregularly aligned tips.} The height over which the fluctuations decay within the canopy and the magnitude of the fluctuations in the core of the canopy appear to correlate with the element spacing. Note that this plateau in $u'$ and $w'$ within the canopy is asymptotic and requires a sufficiently large canopy depth to occur. Thus, this plateau is essentially absent for the canopy of S48, because of its low canopy height-to-spacing ratio, $h/s \approx 2$. The wall-normal fluctuations within the canopy do not exhibit this plateau and decay gradually below the canopy-tip plane to meet the impermeability condition at the canopy base. {Let us also note here that the element-induced flow accounts for less than $30\%$ of the magnitude of the streamwise velocity fluctuations and less than $10\%$ of the cross-velocity fluctuations within the canopy, which is consistent with the observations of \citet{Poggi2008}. This implies that the velocity fluctuations deep within the canopy result mainly from the penetration of the overlying, element-incoherent velocity fluctuations. This will be discussed further in \S\ref{subsec:KH}.}

{Although, as discussed in the preceding paragraphs, the element spacing has a leading-order effect on the fluctuating flow, their height, $h$, also plays a secondary role. In order to assess the effect of height, we consider the canopies of family H, which have fixed element width and spacing, but different canopy heights.} As noted previously, the differences between the element-induced fluctuations for the fixed-spacing canopies of family H are negligible. However, we do observe changes in the full rms velocity fluctuations for these cases, {implying that the height affects the element-incoherent flow}. Above the canopy tips, we observe a decrease in $u'$ and an increase in $v'$ and $w'$ with increasing canopy height, similar to the effect of increasing element spacing, as shown in figures~\ref{fig:stats_all}($b,e,h$). Within the canopy, $u'$ and $w'$ for all the cases collapse to the same curves, only departing to meet the no-slip condition at the canopy base. The corresponding magnitude of $v'$ within the canopy, however, increases with canopy height up to $h/s \approx 6$, {and saturates for $h/s \gtrsim 6$. This saturation is also observed for the effect of the canopy on the flow in general as} illustrated in figures~\ref{fig:stats_all}($b,e,h$) and \ref{fig:U_uv}($b,e$), which show that the velocity fluctuations, Reynolds shear stresses and the mean velocity profiles for cases H96 and H128, with $h/s \approx 6$--$8$ are essentially the same. {The changes in the element-incoherent flow observed for different element heights likely result from a modulation of the Kelvin--Helmholtz-like instability, discussed in \S\ref{subsec:KH}, which is essentially independent of the element-induced flow.}

\begin{figure}
\centering
		 \includegraphics[scale=0.99,trim={10.2mm 3mm 0mm 0mm},clip]{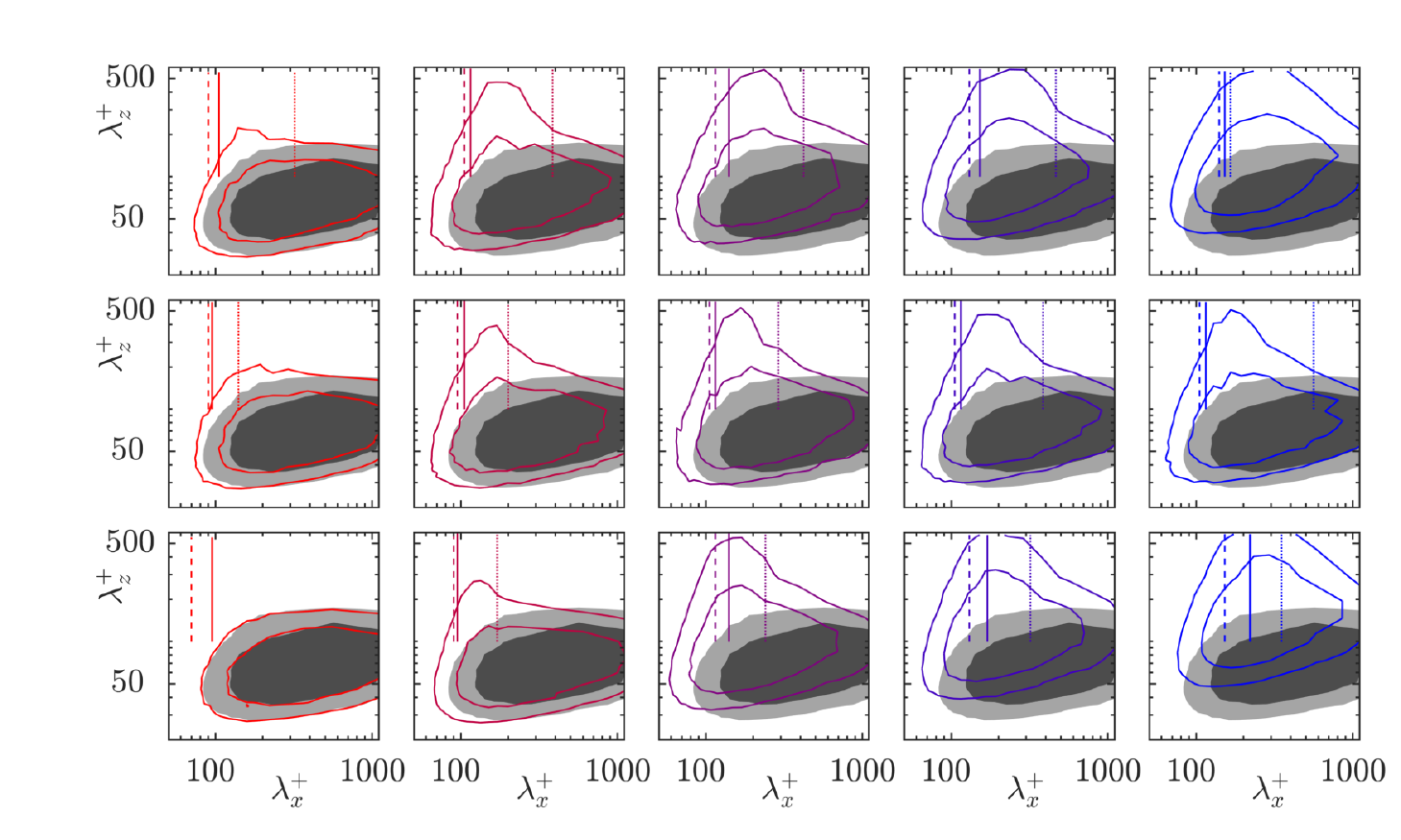}%
         \mylab{-11.2cm}{7.5cm}{(\textit{a})}%
  		\mylab{-8.6cm}{7.5cm}{(\textit{b})}%
  		\mylab{-6.05cm}{7.5cm}{(\textit{c})}%
  		\mylab{-3.5cm}{7.5cm}{(\textit{d})}%
  		\mylab{-0.9cm}{7.5cm}{(\textit{e})}%
  		\mylab{-11.2cm}{5.05cm}{(\textit{f})}%
  		\mylab{-8.6cm}{5.05cm}{(\textit{g})}%
  		\mylab{-6.05cm}{5.05cm}{(\textit{h})}%
  		\mylab{-3.5cm}{5.05cm}{(\textit{i})}%
  		\mylab{-0.9cm}{5.05cm}{(\textit{j})}%
  		 \mylab{-11.2cm}{2.60cm}{(\textit{k})}%
  		\mylab{-8.6cm}{2.60cm}{(\textit{l})}%
  		\mylab{-6.15cm}{2.60cm}{(\textit{m})}%
  		\mylab{-3.5cm}{2.60cm}{(\textit{n})}%
  		\mylab{-0.9cm}{2.60cm}{(\textit{o})}%
  		\caption{Pre-multiplied spectral energy densities of the wall-normal velocity, $k_x k_z E_{vv}$, at height $y^+ \approx 15$, normalised by their respective rms values. The line contours represent ($a$--$e$) cases S10 to S48; ($f$--$j$) cases H16 to H128; and ($k$--$o$) cases G10 to G100. The shaded contours represent the smooth-wall case, SC. The contours are in increments of $0.06$ for all the cases. The vertical lines mark the most amplified wavelength predicted by linear stability analysis, discussed in \S\ref{sec:stability_analysis}; \protect\blackline, DNS mean profiles without drag on fluctuations; \protect\blackdotted, DNS mean profiles with drag on fluctuations; \protect\blackdashed, synthesised mean profiles without drag on fluctuations.}%
   	     \label{fig:spectra_h15_v_all}
\end{figure}

\subsection{{Effect of canopy parameters on the shear-layer instability}}\label{subsec:KH}
The variations observed in the velocity fluctuations for the fixed-element-spacing simulations, discussed above, may result from the growth of the Kelvin--Helmholtz-like, shear-layer instability, typically reported in dense canopy flows \citep{Finnigan2000,Nepf2012}. In order to assess the presence of this instability in the flow, we compare the pre-multiplied spectral energy densities of the wall-normal velocity at $y^+ \approx 15$ in figures~\ref{fig:spectra_h15_v_all}($f$--$j$). For case H16 we observe that the spectral energy densities of the fluctuations above the canopy are similar to those above smooth walls. As the height of the canopy is increased, we observe a progressive increase in the energy in long spanwise wavelengths, $\lambda_z^+ > 100$, for a narrow range of streamwise wavelengths, $\lambda_x^+ \approx 150$--$250$. This range of streamwise wavelengths remains roughly constant for increasing canopy heights. Such a signature in the spectral energy densities has been previously associated with the presence of spanwise-coherent, Kelvin--Helmholtz-like instabilities over riblets \citep{Garcia-Mayoral2011}, transitional roughness \citep{Abderrahaman2019} and permeable substrates \citep{GG2019}. {This signature in the spectral energy densities is also reflected in the instantaneous realisations of the wall-normal velocity, portrayed in figure~\ref{fig:v_snap_G}, which show increased spanwise coherence with increasing element height.} The shear-layer instability is known to generate strong wall-normal fluctuations and, hence, its signature is most clear in the wall-normal spectra \citep{Garcia-Mayoral2011,GG2019}. {\citet{Ghisalberti2009} and \citet{Nepf2012} concluded that canopies only exhibit a shear-layer instability if their height is larger than the wall-normal extent of the instability as otherwise the rollers would be constrained by the lack of canopy depth. Above a short canopy, like that of H16, the proximity of the impermeability condition at the base of the canopy would inhibit the instability by blocking the wall-normal fluctuations. Similarly, \citet{Huerre1983} and \citet{Healey2009} showed that the confinement of a free-shear layer also results in stabilisation of the associated Kelvin--Helmholtz instability.} Increasing the canopy height weakens this effect, leading to a stronger signature of the instability, observed in figures~\ref{fig:spectra_h15_v_all}($f$--$j$). {This enhanced signature of the instability is likely responsible for the increase in the cross-velocity fluctuations for the canopies family H with increasing height, discussed in \S\ref{subsec:fluc}. For $h/s > 6$, the instability no longer perceives the canopy base and the effect of the height on the instability and, consequently, the velocity fluctuations, saturates.} The Kelvin--Helmholtz-like instability has also been reported to cause an increase in the Reynolds shear stresses, with an associated increase in the friction drag, over surfaces such as riblets and permeable substrates \citep{Garcia-Mayoral2011, GG2019}. The increase {and saturation of the Reynolds shear stresses with increasing canopy height can be observed in figure~\ref{fig:U_uv}($e$) for the canopies of family H, and is concurrent with the effect of the element height on the instability, discussed above. This trend in the Reynolds shear stress has a corresponding effect on the drag exerted on the overlying flow, which is illustrated by the downward shift in the mean velocity profiles, portrayed in figure~\ref{fig:U_uv}($b$). The above discussion suggests that the secondary effect that the height has on the full velocity fluctuations within and above the canopy is mainly through its influence on the Kelvin--Helmholtz-like instability.}

\begin{figure}
\centering
		 \vspace{2mm}\includegraphics[scale=1.0,trim={8mm 6mm 0mm 0mm},clip]{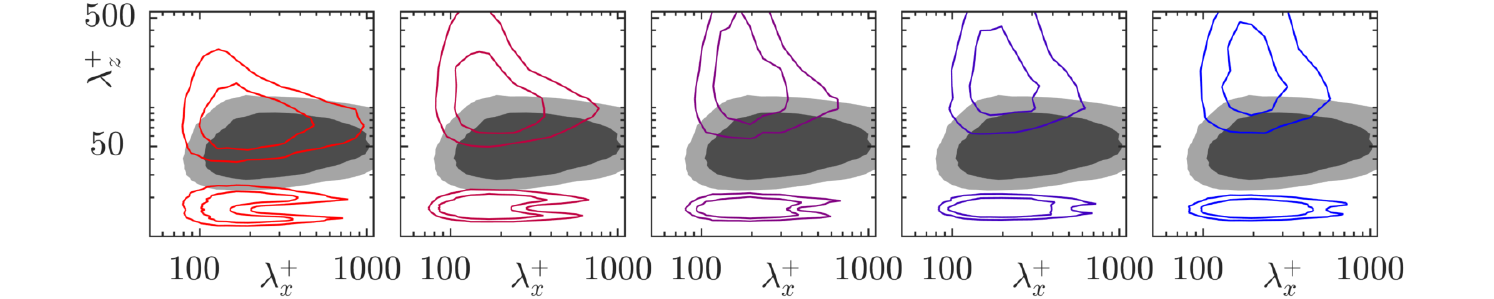}%
         \mylab{-13.6cm}{2.0cm}{(\textit{a})}%
  		\mylab{-11.1cm}{2.0cm}{(\textit{b})}%
  		\mylab{-8.5cm}{2.0cm}{(\textit{c})}%
  		\mylab{-6.0cm}{2.0cm}{(\textit{d})}%
  		\mylab{-3.4cm}{2.0cm}{(\textit{e})}%
  	   \vspace{1mm} \includegraphics[scale=1.0,trim={8mm 0mm 0mm 0mm},clip]{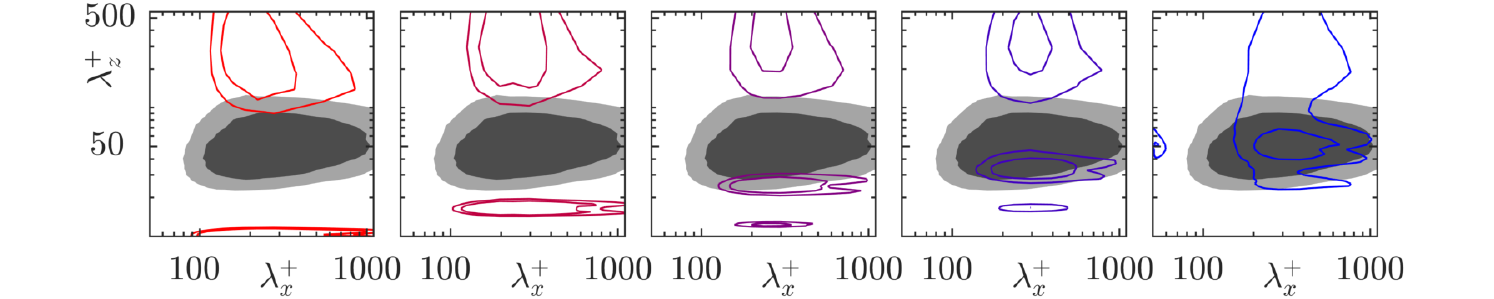}%
         \mylab{-13.6cm}{2.6cm}{(\textit{f})}%
  		\mylab{-11.1cm}{2.6cm}{(\textit{g})}%
  		\mylab{-8.5cm}{2.6cm}{(\textit{h})}%
  		\mylab{-6.0cm}{2.6cm}{(\textit{i})}%
  		\mylab{-3.4cm}{2.6cm}{(\textit{j})}%
  		\caption{Pre-multiplied spectral energy densities of the wall-normal velocity, $k_x k_z E_{vv}$, for ($a$--$e$) cases H16 to H128 at a height of $y^+ \approx -10$; and ($f$--$j$) cases S10 to S48 at a height of $y^+ \approx -40$. The contours are normalised by the rms values of their respective cases. The shaded contours are of the smooth-wall case, SC at a height of $y^+ \approx 1$, for reference. The contours are in increments of 0.075 for all the cases.}%
   	     \label{fig:spectra_H_can_-10}
\end{figure}

\begin{figure}
	\centering
  		\vspace{2mm}\subfloat{%
  	    \includegraphics[scale=1,trim={20mm 39mm 18mm 22mm},clip]{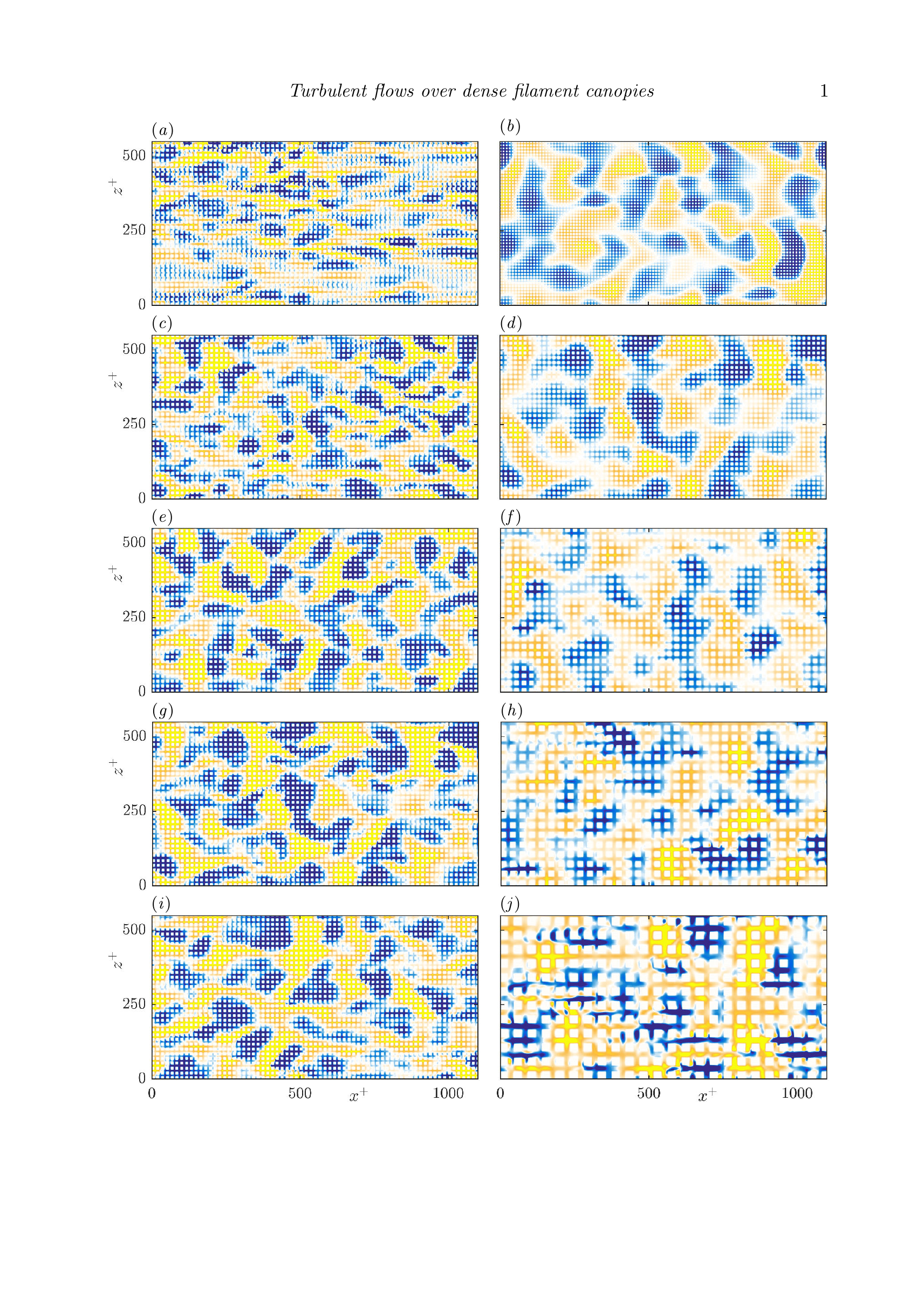}
 		}%
 		\caption{Instantaneous realisations of the wall-normal velocity at $y^+ = -10$ (left column) and $y^+ = -40$ (right column), normalised by $u_\tau$. From top to bottom, the left column represents cases H16 to H128; and right column, cases S10 to S48. From top to bottom, the clearest and darkest colours indicate intensities of $\pm(0.1,0.2,0.3,0.3,0.3)$ in the left column and $\pm(0.05,0.2,0.4,0.4,0.5)$ in the right column.}%
   	     \label{fig:v_snap_H_S}
\end{figure}

The increase in intensity of the Kelvin--Helmholtz-like rollers with increasing canopy height also contributes to the increase in the wall-normal velocity fluctuations within the canopy observed in figure~\ref{fig:stats_all}($e$). This is demonstrated by the wall-normal spectral energy densities of the flow within the canopies, portrayed at $y^+ \approx -10$ for all the canopies of family H, in figures~\ref{fig:spectra_H_can_-10}($a$--$e$). Note that in calculating the spectra for a region with solid obstacles, we have implicitly assumed that the obstacles are fluid regions with zero flow velocity. As discussed in the previous paragraph, for case H16 the instability is inhibited by the proximity of the canopy-base wall, and the flow above shows similarities to a smooth-wall flow. The energy density within the canopy at $y^+\approx -10$ for this case also shows some regions overlapping with the smooth-wall spectra, with additional energy in the wavelengths associated with the Kelvin--Helmholtz-like instability. The smooth-wall spectra displayed for reference are at $y^+ \approx 1$, which is as low as possible while yielding a non-negligible energy, since no direct comparison with $y^+ \approx -10$ is possible. We also observe some energy in the spanwise wavelength corresponding to the canopy spacing and a broad range of streamwise wavelengths. These regions can be attributed to the modulation of the element-induced flow by the larger scale fluctuations induced by the instability or the overlying turbulence \citep{Abderrahaman2019}. This suggests that the fluctuations within a short canopy result mainly from the penetration of the overlying turbulence, with additional contributions from the Kelvin--Helmholtz-like instability and the element-induced flow. As the canopy height is increased, and the instability becomes stronger, the deviations in the spectral energy densities from smooth-wall flow become more prominent. The fluctuations within the canopy in cases H32 to H128 arise mainly from large spanwise wavelengths, likely originating from the Kelvin--Helmholtz-like instability near the canopy-tip plane, along with a contribution of the modulated element-induced flow discussed above. The increasing signature of the instability within the canopy with increasing element height can also be observed in the instantaneous realisations of the wall-normal velocity at $y^+ \approx -10$ portrayed in figure~\ref{fig:v_snap_H_S}. The presence of large spanwise wavelengths deep within the canopy can also be noted for the canopies of family S, whose spectral energy densities and realisations of wall-normal velocity  at $y^+ \approx -40$ are portrayed in figures~\ref{fig:spectra_H_can_-10}($f$--$j$) and \ref{fig:v_snap_H_S}, respectively. {This suggests that, in the present dense canopies, the background turbulence is not able to penetrate far below the canopy tips, and that the velocity fluctuations deep within originate mainly from the footprint of the Kelvin--Helmholtz-like rollers above}.

It is also worth noting that even in canopies with small element spacings, such as that of case S10, although the {fluctuations of the wall-parallel velocities decay rapidly below the canopy-tip plane, the fluctuations of the wall-normal velocity decay more slowly,} as shown in figure~\ref{fig:stats_all}. This is also the case for the velocity fluctuations of the canopies of family H, for which the wall-normal velocity fluctuations within the canopy require larger canopy heights to saturate compared to the wall-parallel fluctuations. The presence of wall-normal fluctuations deep within the canopy are a reflection of the canopy layout being able to obstruct the wall-normal flow less efficiently than the tangential flow. It will be shown in \S\ref{sec:stability_analysis} that, for the present canopies, the effective drag coefficient in the tangential directions can be up to three times larger than in the wall-normal direction. In the core region of a tall canopy, the only mechanism to inhibit the velocity fluctuations is the canopy drag. As the canopy geometries studied here exert more drag on the wall-parallel flow than the wall-normal flow, $u'$ and $w'$ decay faster than $v'$ within the canopy.

{The spacing between the canopy elements affects the excitation of the Kelvin--Helmholtz-like instability through its influence on the canopy drag. \citet{White2007} and \citet{AS2017} have shown that the canopy drag governs the instability through two competing effects, the shear at the canopy tips and the canopy drag. A small element spacing results in a large drag within the canopy, which in turn results in a larger shear at the canopy tips that enhances the instability, but at the same time it also inhibits the fluctuations more strongly, which weakens the instability.} To study this effect, we now compare the pre-multiplied spectral energy densities of the wall-normal velocities for the canopies of family S, which have a constant height and different element spacing. For the canopy with the smallest spacing, S10, {the signature of the Kelvin--Helmholtz-like instability in the energy densities is weak and the distribution of energy in different wavelengths is similar to that over smooth walls, as shown in figure~\ref{fig:spectra_h15_v_all}($a$). This suggests that the  large drag exerted on the fluctuations by this canopy inhibits the formation of the shear-layer instability. As the element spacing is increased, the drag on the fluctuations reduces, and there is a progressive increase in the energy in wavelengths associated with the Kelvin--Helmholtz-like instability for the canopies of S16--S48, as can be observed in figures~\ref{fig:spectra_h15_v_all}($b$--$e$). The change in element spacing also has an effect on the instability wavelength, which will be discussed later in this section.} In addition to the increase in the energy associated with the instabilities, the increase in element spacing also results in a progressive decrease in the overlapping regions in the energy densities of the canopy and smooth-wall flows, with a reduction in the energy in wavelengths $\lambda_x^+ \gtrsim 700$, {$\lambda_z^+ \approx 50$--$100$}. If the element spacing was increased further, the Kelvin--Helmholtz-like instability would eventually weaken, and for sparse enough canopies the flow within would begin to resemble smooth-wall flow perturbed by the element-induced flow of the isolated canopy elements. {Such sparse canopies are beyond the scope of the present work, but have been discussed in \citet{AS2019}, as well as in the previous studies of \citet{Poggi2004, Pietri2009} and \citet{Huang2009}.}

\begin{figure}
\centering
		 \includegraphics[scale=1.0,trim={6mm 5mm 0mm 8mm},clip]{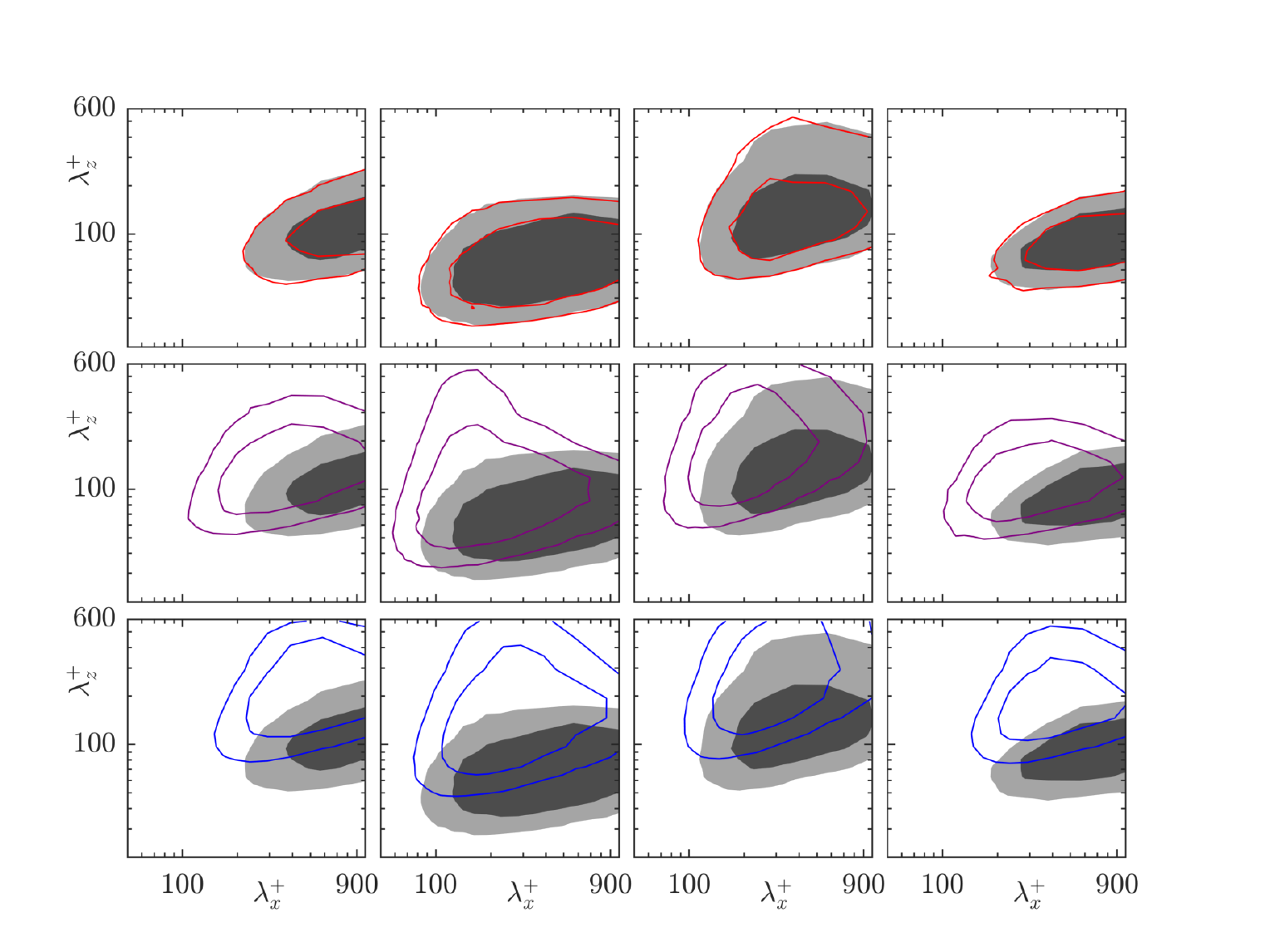}%
         \mylab{-13.9cm}{9.45cm}{(\textit{a})}%
  		\mylab{-10.8cm}{9.45cm}{(\textit{b})}%
  		\mylab{-7.7cm}{9.45cm}{(\textit{c})}%
  		\mylab{-4.6cm}{9.45cm}{(\textit{d})}%
  		\mylab{-13.9cm}{6.35cm}{(\textit{e})}%
  		\mylab{-10.8cm}{6.35cm}{(\textit{f})}%
  		\mylab{-7.7cm}{6.35cm}{(\textit{g})}%
  		\mylab{-4.6cm}{6.35cm}{(\textit{h})}%
  		\mylab{-13.9cm}{3.2cm}{(\textit{i})}%
  		\mylab{-10.8cm}{3.2cm}{(\textit{j})}%
  		\mylab{-7.7cm}{3.2cm}{(\textit{k})}%
  		\mylab{-4.6cm}{3.2cm}{(\textit{l})}%
  		\mylab{-12.9cm}{9.9cm}{$k_x k_z E_{uu}$}%
   		 \mylab{-9.8cm}{9.9cm}{$k_x k_z E_{vv}$}%
   		 \mylab{-6.80cm}{9.9cm}{$k_x k_z E_{ww}$}% 	
   		 \mylab{-3.80cm}{9.9cm}{$k_x k_z E_{uv}$}% 	
  		\caption{Pre-multiplied spectral energy densities at $y^+ \approx 15$ normalised by their respective rms value. The line contours represent ($a$--$d$) case G10; ($e$--$h$) case G40; ($i$--$l$) case G100. The filled contours represent the smooth-wall case, SC. The contours in ($a$, $e$, $i$), ($b$, $f$, $j$), ($c$, $g$, $k$) and ($d$, $h$, $l$) are in increments of $0.075$, $0.06$, $0.07$, and $0.1$, respectively.}%
   	     \label{fig:spectra_h15_G}
\end{figure}

Let us now focus on the self-similar canopy geometries of family G. {Although these canopies have the same $\lambda_f$, increasing the size of the canopy elements produces similar effects on the pre-multiplied spectral energy densities as the canopies of family S, discussed in the previous paragraph.} For the densest canopy, G10, the spectral energy densities at $y^+ \approx 15$  collapse with those over a smooth wall, as shown in figures~\ref{fig:spectra_h15_G}($a$--$d$). As the size of the canopy is increased, we observe a stronger signature of the Kelvin--Helmholtz-like instability in the energy densities, {portrayed in figures~\ref{fig:spectra_h15_v_all}($k$--$o$). The associated increase in spanwise coherence in the flow can also be observed in the instantaneous realisations of the wall-normal velocity shown in figure~\ref{fig:v_snap_G}. Note that here, we observe the combined effects of increasing canopy height, as in family H, and increasing spacing, as in family S, on the instability. So far we have mainly discussed the spectral energy densities of the wall-normal velocity fluctuations, as they have the strongest signature of the Kelvin--Helmholtz-like instability. For completeness, we now} use the canopies of family G to illustrate the effect of increasing the canopy size on the spectral energy densities of the streamwise and spanwise fluctuations, and the Reynolds shear stresses, portrayed in figure~\ref{fig:spectra_h15_G}. The distinct region in the wall-normal spectral energy densities associated with the Kelvin--Helmholtz-like instability is not so apparent in the energy densities of the other velocity fluctuations and the Reynolds shear stresses. Nevertheless, as the canopy size increases, we observe an increase in the energy in streamwise wavelengths associated with the instability, along with a general increase in the energy in {shorter and wider} wavelengths compared to smooth-wall flows. {This is consistent with the gradual shortening and widening of the eddies observed in figure~\ref{fig:v_snap_G}.}

The results discussed in this section suggest that the growth of the Kelvin--Helmholtz-like instability depends on both the canopy height and the element spacing. The streamwise wavelength of the instability, however, seems to depend mainly on the element spacing. {We observe that, in the canopies of family H the streamwise wavelength of the instability is roughly constant regardless of the canopy height, $\lambda_x^+ \approx 150$, as can be observed in figures~\ref{fig:spectra_h15_v_all}($f$--$j$). For the canopies of family S, however,  the increase in the element spacing results in an increase in the streamwise wavelength of the instability from $\lambda_x^+ \approx 140$ for case S10 to  $\lambda_x^+ \approx 280$ for case S48, as shown in figures~\ref{fig:spectra_h15_v_all}($a$--$e$). Similarly, for the canopies of family G there is a progressive increase in the streamwise wavelengths associated with the instability with the increase in canopy size.} The streamwise wavelength of the Kelvin--Helmholtz-like instability is determined by the shear length, {typically defined in the literature as $L_s = U/(dU/dy)$ calculated at the canopy-tip plane \citep{Raupach1996,Finnigan2000,Nepf2012}. Previous studies have shown that the shear length $L_s$ in canopy flows can be determined by the effective streamwise canopy drag coefficient \citep{Finnigan2000,Nepf2007,Ghisalberti2009,Nepf2012}. Intuitively, in tall, dense canopies, we would expect this drag coefficient, and by extension the shear-layer thickness, to be a function of the element spacing. A dependence of the canopy shear-layer thickness on the element spacing was also observed by \citet{Novak2000} in their study of natural canopy flows.} Therefore, the canopies of family H, which have a constant element spacing, have similar mean drag coefficients and, consequently, similar instability wavelengths, as observed in the spectral energy densities of the fixed-spacing canopies. For the canopies of families S and G, increasing the element spacing decreases the canopy drag coefficient, thereby resulting in the larger wavelengths observed {for the shear-layer eddies}. The effect of the canopy spacing on the drag coefficients and the instability wavelengths will be discussed further in \S\ref{sec:stability_analysis}. 

%\subsubsection{Effect of $\Rey_\tau$ on the Kelvin--Helmholtz-like instability}
{Finally, let us discuss the effect of the Reynolds number on the Kelvin--Helmholtz-like instability.} It was shown in \S\ref{sec:Rey_effects}, that the turbulent fluctuations over dense canopies scale in friction units, and therefore similar results are obtained when simulating canopies with the same height and spacing in friction units at different Reynolds numbers. It can be observed in figure~\ref{fig:spectra_h32_Re_var}($b$) that the signature of this instability in simulations H32$_{180}$ and H32$_{400}$ are essentially the same, and that the associated streamwise wavelength for both cases is roughly $\lambda_x^+ \approx 150$.  {As discussed above, the wavelength and amplification of the instability are governed by the shear at the canopy tips. As the canopy parameters for cases  H32$_{180}$ and H32$_{400}$ are kept constant in friction units, we can also expect the shear at the canopy tips to also be similar. Therefore, the instability characteristics for both these canopies are essentially the same when scaled in friction units. We have observed similar behaviours for Kelvin--Helmholtz-like instabilities originating over riblets \citep{Garcia-Mayoral2012} and permeable substrates \citep{GG2019}.}

\section{Linear analysis of Kelvin--Helmholtz-like instabilities}\label{sec:stability_analysis}
The results from DNS discussed in \S\ref{sec:turb_fluct} show that the flow in the region near the canopy-tip plane can be dominated by the presence of spanwise-coherent structures originating from a Kelvin--Helmholtz-like instability. This instability can be captured by a two-dimensional, mean-flow linear stability analysis, even in turbulent flows \citep{Jimenez2001,White2007,Garcia-Mayoral2011,Zampogna2016,GG2019}. In this section, we discuss the methodology and results from such an analysis conducted on the velocity profiles obtained from the DNSs. As the Kelvin--Helmholtz-like instability is an inviscid phenomenon, several of the studies just cited use an inviscid analysis to capture it. The inclusion of viscosity, however, inhibits the growth of smaller wavelengths in the flow, and consequently, results in the most amplified wavelength being slightly larger compared to that of an inviscid analysis \citep{Jimenez2001,GG2019}. In this section, we present the results only from viscous analysis. The results from an inviscid analysis are presented in appendix~\ref{appB} for reference. In addition, we show that some of the key features of this instability can be captured by linear analysis performed on velocity profiles modelled a priori, which would not require any information from the DNSs. 

For the purpose of the stability analysis, we model the effect of the canopy using a drag force in the Navier--Stokes equations, which results in the following governing equations
\begin{eqnarray}
\frac{\p \mathbf{u}}{\p t} + \mathbf{u}\bcdot\nabla\mathbf{u} & = & -\nabla p + {\nu} \nabla^2 \mathbf{u} - \nu {C_i} \mathbf{u}\label{eq:NS_plus_drag}\\
\nabla\bcdot\mathbf{u} & = & 0,\label{eq:cont}
\end{eqnarray}
where $C_i$ is the effective canopy drag coefficient in each $i^{th}$ direction, {has dimensions of inverse length squared --being essentially the inverse of a permeability--, and is} assumed to be homogeneous over the entire canopy region, as in \citet{Singh2016}. Given the density of the canopies considered, with maximum spacings $s^+ = \mathcal{O}(10)$, we assume that inertial effects in the flow deep within the canopy are small and can be neglected. {In addition, the element width of the canopies is also small, $w^+ \approx 1$--$25$. For such canopies, the canopy drag can be assumed to depend linearly on the velocity \citep{Tanino2005,Tanino2008}.}

\begin{figure}
	\centering
		\vspace{0mm}\subfloat{%
%  		 \tikzsetnextfilename{cd_vs_w}
  		\hspace{0mm}\includegraphics[scale=1]{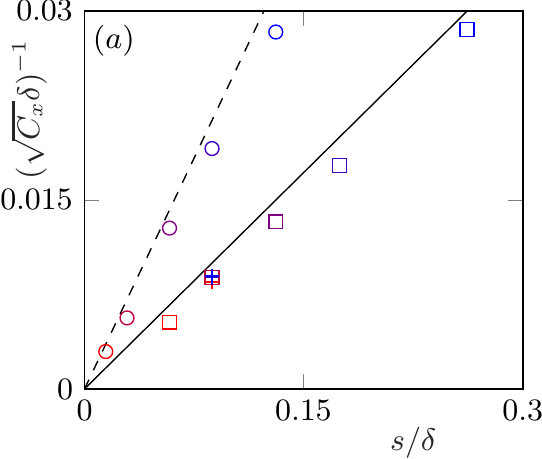}
  		}%
  		\subfloat{%
 % 		 \tikzsetnextfilename{cy_vs_w}
  			\hspace{4mm}\includegraphics[scale=1]{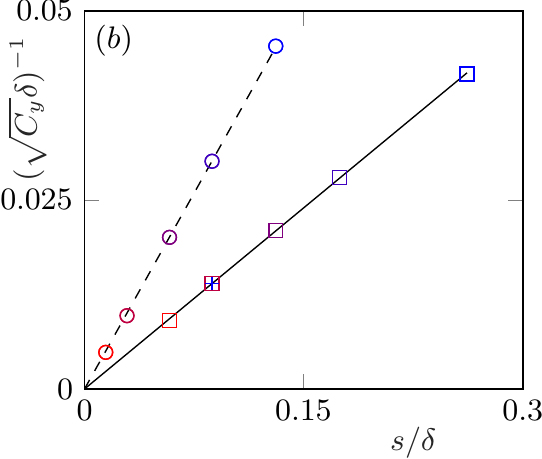}
  		}%
 		\caption{Variation of the lengthscales derived from the ($a$) streamwise and ($b$) wall-normal canopy drag coefficients for different element spacings. The symbols represent, \protect\hsquare, cases of S; $+$, cases of H; and \protect\hcircle, cases of G. The colours from red to blue represent cases S10 to S48, H16 to H128 and G10 to G100. {The symbols in ($a$) are values obtained from the DNSs, and the dashed and solid lines are predictions from two-dimensional Stokes-flow simulations. Both the symbols and the lines in ($b$) are obtained from Stokes-flow simulations.}}
  		 \label{fig:cd_vs_s}
\end{figure}

In the core of the canopy, away from the shear effects at the canopy base and top, the mean momentum equation would reduce to a balance between the canopy drag and the mean pressure gradient
\begin{equation}
{\nu C_x U = -\frac{\mathrm{d}P}{\mathrm{d} x}.}
\label{eq:canopy_drag_balance}
\end{equation}
Equation~\eqref{eq:canopy_drag_balance} is essentially Darcy's equation for flow within permeable substrates \citep{Darcy1856}, and has been used by \citet{Zampogna2016a} to model flow deep within densely packed, rigid fibres. The streamwise drag coefficient, $C_x$,  can be obtained by substituting the values of $U$ and ${\mathrm{d}P}/{\mathrm{d}x}$ obtained from the DNSs into equation~\eqref{eq:canopy_drag_balance}. From dimensional arguments, equation~\eqref{eq:canopy_drag_balance} predicts that the drag coefficient would scale as $C_x \sim 1/s^2$. This scaling is demonstrated in figure~\ref{fig:cd_vs_s}($a$), which suggests that equation~\eqref{eq:canopy_drag_balance} provides a reasonable approximation for the flow deep within the present canopies, excluding the sparsest canopy S48. Although we can expect the flow within the canopy to be Darcy-like in the wall-normal direction as well, we cannot use the DNS results to obtain $C_y$, as there is no mean flow in this direction. In order to obtain $C_y$, we consider separately the Stokes flow along infinitely long canopy elements driven by a constant pressure gradient. The equation for such flow is $\nu (\p^2_x + \p^2_z)v = \mathrm{d}P/\mathrm{d}y$. The wall-normal drag coefficient is then obtained as
\begin{equation}
{\nu C_y \langle v \rangle = -\frac{\mathrm{d}P}{\mathrm{d}y}, }
\label{eq:wall_normal_drag}
\end{equation}
where the angled brackets represent a spatial average. The estimated values of $C_y$ are portrayed in figure~\ref{fig:cd_vs_s}($b$) for reference. It may be noted that the ratio of the streamwise to wall-normal drag coefficients for the present canopies is $C_x/C_y \approx 2$--$3$, which shows that the streamwise flow is more obstructed than the wall-normal flow for the layouts considered. {It is worth noting here that the canopy drag coefficients, $C_x$ and $C_y$, and the ratio between them also depends on significantly on the plan view arrangement and the resulting porosity of the canopy \citep{VanderWest1996,Zampogna2016a}. This is evidenced by the different ratios of the drag lengthscale and the element spacing for the canopies of families S and G, portrayed in figure~\ref{fig:cd_vs_s}, which have width-to-spacing ratios, $w/s = 1/2$ and $2/9$, respectively. The dependence of the $C_x$ on $w/s$ can also be predicted using two-dimensional Stokes flow simulations, as shown in figure~\ref{fig:cd_vs_s}($a$). For more complicated canopy arrangements, such as staggered or random, we would expect the drag to be a function of the planar layout of the elements.}

{In order to conduct the stability analysis, we linearise} the equations~\eqref{eq:NS_plus_drag} and \eqref{eq:cont} around the mean flow, $U(y)$, {yielding}
\begin{eqnarray}
\frac{\p {u}}{\p t} + U \frac{\p u}{\p x} + v U' & = & -\frac{\p p}{\p x} + {\nu} \nabla^2 u - \nu {C_x} u\\
\frac{\p {v}}{\p t} + U \frac{\p v}{\p x} & = & -\frac{\p p}{\p y} + {\nu} \nabla^2 v - \nu {C_y} v\\
\frac{\p u}{\p x}  +  \frac{\p v}{\p y} & = & 0.
\label{eq:NS_plus_drag_lin}
\end{eqnarray}
These equations are used to obtain a modified Orr-Sommerfeld equation \citep{Drazin2004,White2007,Singh2016,Zampogna2016},
\begin{eqnarray}
\left(\frac{\p}{\p t} + U \frac{\p}{\p x} + \nu C_y \right)\nabla^2 v - {\nu} \nabla^4 v = U'' \frac{\p v}{\p x} - \nu(C_x - C_y) \frac{\p^2 v}{\p y^2}.
\label{eq:NS_lin_v}
\end{eqnarray}
Assuming wavelike solutions of the form $v = \widetilde{v} e^{i (\alpha x - \omega t)}$, equation~\eqref{eq:NS_lin_v} reduces to the eigenvalue problem
\begin{multline}
(\alpha U - i \nu C_y)(\mathrm{D}^2   -  \alpha^2)\widetilde{v} 
- \alpha U'' \widetilde{v}  - i\nu(C_x - C_y) D^2 \widetilde{v} \\   + {i\nu}(\mathrm{D}^4 - 2\alpha^2 \mathrm{D}^2 + \alpha^4)\widetilde{v}= \omega (\mathrm{D}^2 - \alpha^2) \widetilde{v},
\label{eq:OS}
\end{multline}
where the prime superscript denotes differentiation with respect to $y$, and $\mathrm{D}$ represents the operator $\mathrm{d}/\mathrm{d}y$. Equation~\eqref{eq:OS} is solved to obtain the complex frequency, $\omega$, for real values of the streamwise wavenumber, $\alpha$, subject to no-slip and impermeability boundary conditions at the top and bottom walls. The instability is then amplified for positive values of the imaginary part of $\omega$.

\begin{figure}
	\centering
		\subfloat{%
%  		 \tikzsetnextfilename{omega_lx_S}
  		 \hspace{0mm}\includegraphics[scale=1,trim= 0mm 0mm 0mm 0mm, clip]{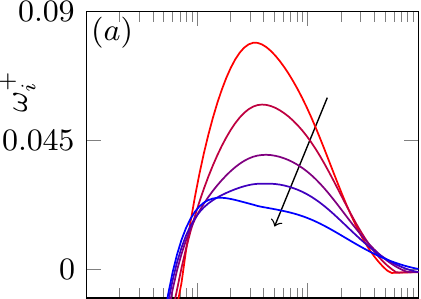}
  		}%
  		\subfloat{%
%  		 \tikzsetnextfilename{omega_lx_H}
  		\hspace{2mm}\includegraphics[scale=1,trim= 0mm 0mm 0mm 0mm, clip]{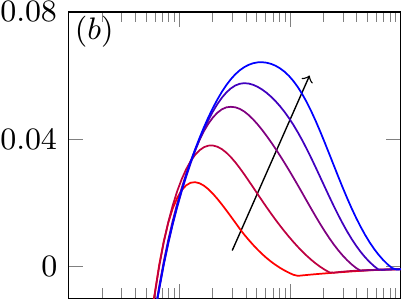}
  		}%
  		\subfloat{%
%  		 \tikzsetnextfilename{omega_lx_G}
  		\hspace{2mm}\includegraphics[scale=1,trim= 0mm 0mm 0mm 0mm, clip]{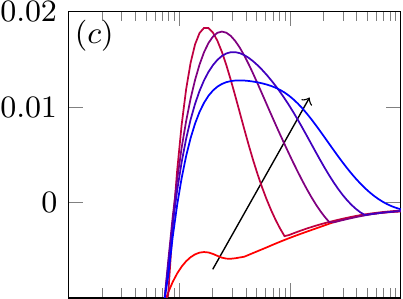}
  		}%
  		
  	\vspace{0mm}\subfloat{%
%  		 \tikzsetnextfilename{omega_lx_S_z}
  		 \hspace{0mm}\includegraphics[scale=1,trim= 0mm 0mm 0mm 0mm, clip]{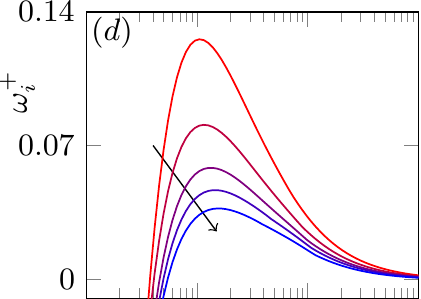}
  		}%
  	    \subfloat{%
%  		 \tikzsetnextfilename{omega_lx_H_z}
  		\hspace{2mm}\includegraphics[scale=1,trim= 0mm 0mm 0mm 0mm, clip]{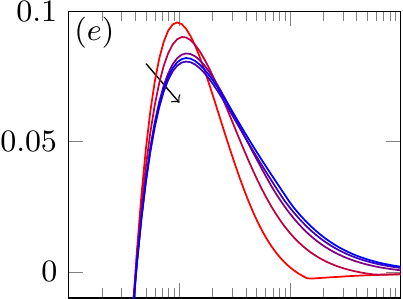}
  		}%
  		\subfloat{%
%  		 \tikzsetnextfilename{omega_lx_G_z}
  		\hspace{2mm}\includegraphics[scale=1,trim= 0mm 0mm 0mm 0mm, clip]{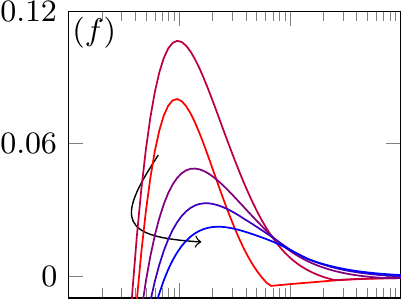}
  		}%
  		
  		 \vspace{0mm}\subfloat{%
%  		 \tikzsetnextfilename{omega_lx_m_S_z}
  		 \hspace{1.6mm}\includegraphics[scale=1,trim= 0mm 0mm 0mm 0mm, clip]{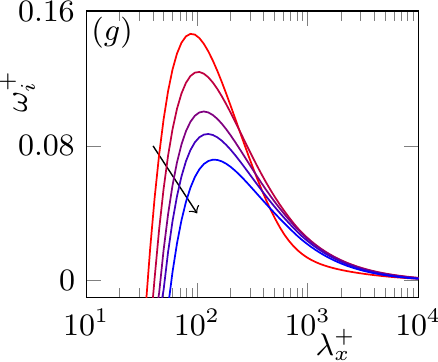}
  		}%
  	    \subfloat{%
%  		 \tikzsetnextfilename{omega_lx_m_H_z}
  		\hspace{0.2mm}\includegraphics[scale=1,trim= 0mm 0mm 0mm 0mm, clip]{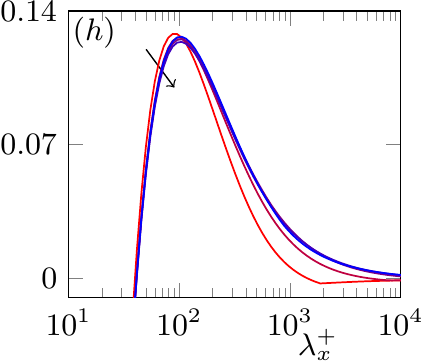}
  		}%
  		 \subfloat{%
%  		 \tikzsetnextfilename{omega_lx_m_G_z}
  		\hspace{0.2mm}\includegraphics[scale=1,trim= 0mm 0mm 0mm 0mm, clip]{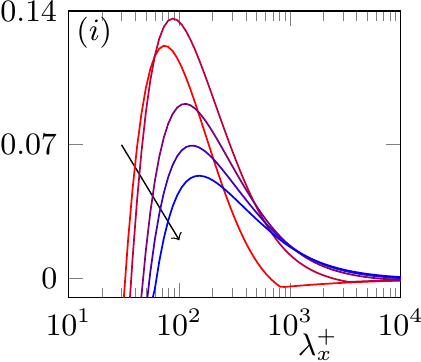}
  		}%
 			 \caption{Growth rates of different perturbation wavelengths obtained from the stability analysis performed on ($a$--$c$) mean profiles obtained from the DNSs, with drag on the perturbations included in the stability analysis; ($d$--$f$) mean profiles obtained from DNSs, with no drag on the perturbations; and ($g$--$i$) mean velocity profiles obtained using equation~\eqref{eq:mean_vel_model}, with no drag on the perturbations. The lines from red to blue, {indicated by the direction of the arrows},  represent ($a$,$d$,$g$) cases S10 to S48; ($b$,$e$,$h$) cases H16 to H128; and ($c$,$f$,$i$) cases G10 to G100.}
 		  	\label{fig:amp_vs_lx}	
\end{figure}

\begin{table}
\begin{center}

%\lineup
\begin{tabular}{clcccc}
   &Case                                                 &DNS     &SA$_{C0}$     &SAM$_{C0}$           & SA \\
\hline
              &S10                                        &140                                  &105          &90       &320                  \\
 \multirow{2}{*}{Fixed}&S16(H96)           &160                                  &115           &105     &385                  \\
  \multirow{2}{*}{height}    &S24             &200                                 &140          &115       &420                  \\
              &S32                                        &230                                 &152         &130      &465                    \\
              &S48                                        &250                                 & 152         &140      &165                    \\
\hline
              &H16                                      &130                                   &95         &90     &140                         \\
\multirow{2}{*}{Fixed}      &H32            &150                                  &105        &95     &200                         \\
\multirow{2}{*}{spacing}   &H64           &160                                  &115        &105    &290                         \\
    &H96(S16)                                         &160                                &115         &105     &385                         \\
            &H128                                        &160                                 &115         &105     &560                         \\
            
\hline
            &G10                                           &--                                   &95          &70     &--                             \\
\multirow{2}{*}{Self-similar}      &G20  &120                               &95          &90      &170                            \\           
\multirow{2}{*}{geometry} 	&G40     &140                                &140        &115     &240                            \\           
									&G60              &190                                &170        &130      &320                           \\           
									&G100            &260                                &220        &152      &350                           \\           
\hline
 \multirow{2}{*}{Varying $\Rey_\tau$} &H32$_{180}$   &140     &105      &--           &180                           \\
                                                                &H32$_{400}$   &140    &105       &--          &180                            \\
\hline
\end{tabular}
\caption{\label{tab:instab_wavelengths} Most amplified instability wavelengths observed in the DNSs and predicted by the stability analysis, scaled in friction units. The column labelled `DNS' lists the approximate streamwise wavelength associated with the instability in the wall-normal spectra portrayed in figure~\ref{fig:spectra_h15_v_all}. SA$_{C0}$, most amplified wavelengths from stability analysis on DNS mean profiles without drag on fluctuations; SAM$_{C0}$, on synthesised velocity profiles without drag on fluctuations; and SA, on DNS mean profiles with drag on fluctuations.}
\end{center}
\end{table} 

\begin{figure}
	\centering
		\subfloat{%
%  		 \tikzsetnextfilename{h96_instab_SF_cd}
  		\includegraphics[scale=1]{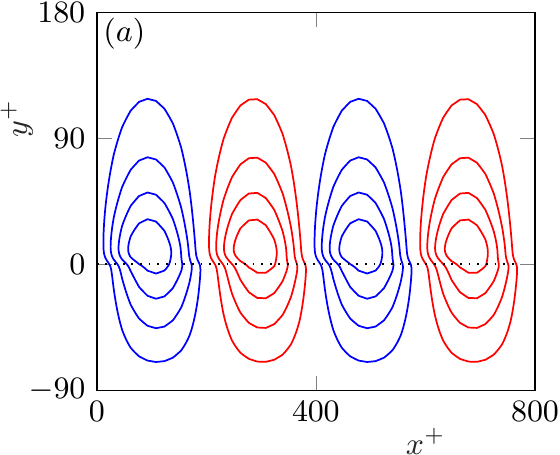}
  		}%
  		\subfloat{%
%  		 \tikzsetnextfilename{h96_instab_SF_cdz}
  		\hspace{7mm}\includegraphics[scale=1,trim=0mm 0mm 0mm -1.1mm, clip]{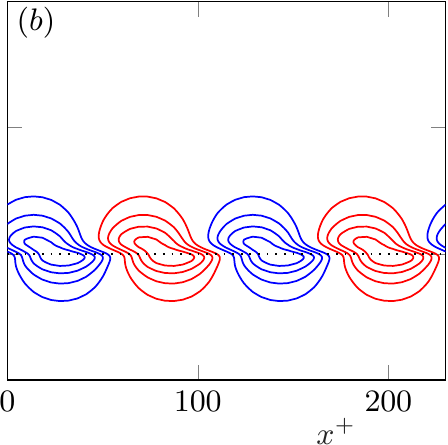}
  		}%
  		  		\caption{Contours of the stream function for the most amplified mode for case H96 obtained from the stability analysis ($a$) with drag and ($b$) without drag on the perturbations. The blue and red lines correspond to clockwise and counter-clockwise rotation, respectively.}
 		  	\label{fig:instab_SF}	
\end{figure}

The growth rates for different perturbation wavelengths are portrayed in figures~\ref{fig:amp_vs_lx}($a$--$c$), and the wavelengths with the highest growth rates are summarised in table~\ref{tab:instab_wavelengths}. The most amplified wavelengths predicted by the stability analysis only match those observed in the DNSs for canopies with high values of $\delta/h$. The wavelengths predicted for cases H16, H32, G10, G20 and G40 show reasonable agreement with those observed in the DNSs. For canopies with larger heights, however, the analysis predicts wavelengths larger than those observed in the DNSs. For the fixed-spacing canopies of family H, the predicted instability wavelength also increases with increasing canopy height, whereas the DNSs show that the instability wavelength for these cases is essentially independent of the height. The contours of the instability stream function for case H96 for the most amplified wavelength, $\lambda_x^+ \approx 385$, portrayed in figure~\ref{fig:instab_SF}($a$), show that it has a large wall-normal span, extending up to $y^+\approx 120$. Such an instability was also reported by \citet{Singh2016}, who performed stability analyses similar to the one conducted here, except that the canopy was represented by a drag force depending quadratically on the velocity. \citet{Singh2016} noted that their analysis predicted two instability modes, one similar to the Kelvin--Helmholtz instability and another originating from the canopy drag included in the analysis. They only considered canopies with low $\delta/h$, and observed that the second instability mode, similar to the large-wavelength modes obtained from the present stability analysis, was dominant for canopies with high drag and spanned the entire height of the channel. 

{It is worth noting here, that in the region near the interface between the canopy and the free-flow, the assumption of a constant drag coefficient given by equation~\eqref{eq:canopy_drag_balance} would no longer be valid, as shear and advective effects become stronger. We have conducted some exploratory analyses accounting for this variation in the drag coefficient and these do not provide improved estimates for the instability wavelength compared to the results of the constant-drag analysis presented here. In the present analysis, we have also assumed} that the drag coefficient experienced by the perturbations is the same as that experienced by the mean flow. {However, we have recently reported for sparser canopies that different wavelengths in the flow can perceive drag coefficients different from that for the mean flow \citep{AS2019}. In such a case, the drag coefficient would have to be calculated on a mode-by-mode basis for the different wavelengths. A wavelength dependent drag coefficient would lead to the drag for a given wavelength being dependent on a convolution from all other wavelengths. Such an analysis, however, is beyond the scope of the present work. An alternative approach would be to model the canopy as a permeable substrate, which naturally yields wavelength-dependent equations for the flow within\citep{Zampogna2016,Abderrahaman2017,AS2017,GG2019}. In order to illustrate that applying constant drag coefficient on the perturbations may be a coarse assumption, we also present results from stability analyses with no drag on the perturbations. This is also a rather coarse assumption, but we observe that excluding} the drag on the perturbations in the stability analysis yields better estimates for the instability wavelengths observed in the DNSs, as shown in figure~\ref{fig:spectra_h15_v_all}. For the canopies of family H, the stability analysis without drag on the fluctuations shows that the most amplified wavelength does not vary significantly with the canopy height. For the canopies of families S and G, this analysis shows an increase in the most amplified wavelength with increasing element spacing, owing to the increase in the shear-layer thickness. The results from this analysis are portrayed in figures~\ref{fig:amp_vs_lx}($d$--$f$), and the most amplified wavelength for each case is listed in table~\ref{tab:instab_wavelengths}. While neglecting the drag acting on the fluctuations yields better estimates for the most amplified wavelengths for canopies with small spacings, the predictions for larger spacings differ by up to a factor of two from the DNS observations. This is likely due to the assumption that the mean flow is homogeneous in the tangential directions, implicit in the stability analysis, which breaks down for such cases. {We have not observed any significant signature of the Kelvin--Helmholtz-like instability for the sparser canopies studied in \citet{AS2019} despite the presence of an inflection in the mean velocity profiles.}  There may also be some distortion of the instability by the ambient turbulent fluctuations in the DNSs \citep{Rogers1994,Raupach1996}.

 \begin{figure}
	\centering
		\subfloat{%
%  		 \tikzsetnextfilename{omega_lx_H_Re}
  		\includegraphics[scale=1]{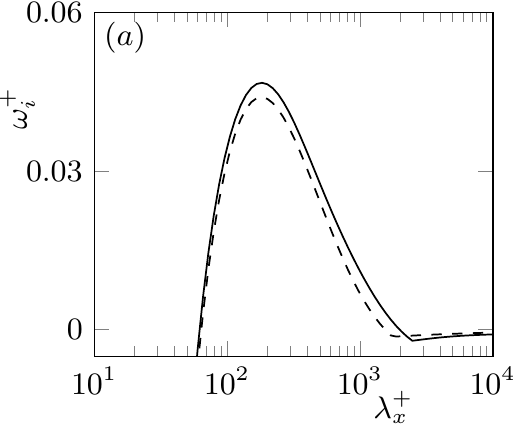}
  		}%
  		\subfloat{%
%  		 \tikzsetnextfilename{omega_lx_H_Re_cdz}
  		\hspace{3mm}\includegraphics[scale=1]{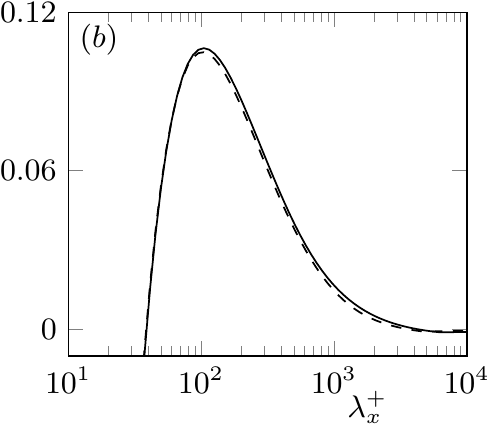}
  		}%
  		\caption{Growth rate for different perturbation wavelengths from the stability analysis for cases \protect\blackline, H32$_{180}$; and \protect\blackdashed, H32$_{400}$; ($a$) with drag  and ($b$) without drag on the perturbations. }
 		  	\label{fig:amp_vs_lx_Re}	
\end{figure}%

{We have also performed stability analyses on the cases with different Reynolds numbers, H32$_{180}$ and H32$_{400}$. These analyses predict similar instability wavelengths and growth rates in viscous units, as shown in figure~\ref{fig:amp_vs_lx_Re}, which is consistent with the observations in the DNSs, discussed in \S\ref{subsec:KH}. These results emphasize that the strength of inflection in the mean-velocity profile and the shear-layer thickness for both these cases is similar when scaled in friction units and, hence, so is their effect on the instability.}

\begin{figure}
	\centering
		\subfloat{%
%  		 \tikzsetnextfilename{Ls_vs_lx}
  		\hspace{-3mm}\includegraphics[scale=1]{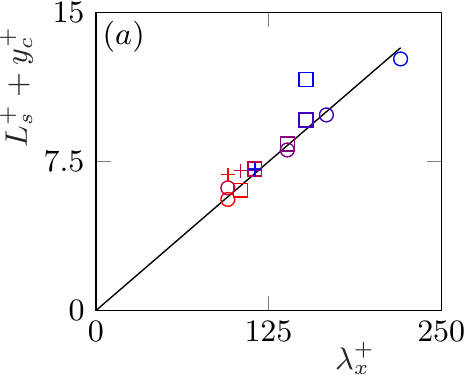}
  		}%
  		\hspace{-0.4mm}\subfloat{%
%  		 \tikzsetnextfilename{cdl_vs_Ls}
  		\includegraphics[scale=1]{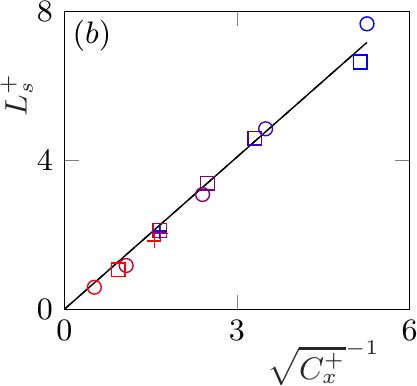}
  		}%
  		\hspace{0mm} \subfloat{%
%  		 \tikzsetnextfilename{s_vs_Ls}
  		\includegraphics[scale=1]{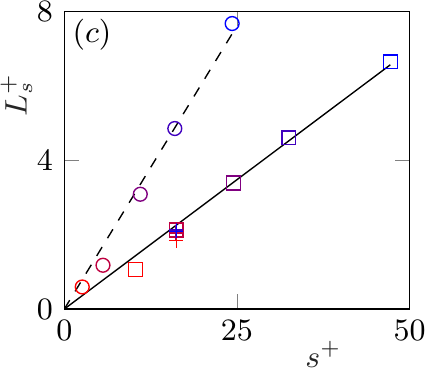}
  		}%
 			 \caption{($a$) Instability wavelength, $\lambda_x^+$, obtained from the linear stability analysis versus the total shear length, $L_s^+ + y_c^+$; ($b$) shear length, $L_s^+$, versus the drag lengthscale; ($c$) shear length versus the element spacing. \protect\hsquare, family S; $+$, family H; \protect\hcircle, family G. The colours from red to blue represent cases S10 to S48, H16 to H128 and G10 to G100. {In ($a$) and ($b$), the solid lines are linear regressions with slopes $0.06$ and $1.36$, respectively. In ($c$), the solid and dashed lines are linear regressions with slopes $0.14$ and $0.3$, respectively.}}
 		  	\label{fig:instab_wavelength_corr}	
\end{figure}

The results obtained from the DNSs and the stability analysis suggest that there is a dependence on the element spacing of the most amplified wavelength, related to the effect of the spacing on the shear-layer thickness. {The usual definition of the shear-layer thickness, $L_s = U/(dU/dy)$,} misses the contribution of the part of the shear layer above the canopy. {Regarding the latter,} \citet{Garcia-Mayoral2011} studied the formation of Kelvin--Helmholtz-like instabilities over riblets, and noted that the shear-layer thickness above was given by the height at which the vorticity gradient, $\mathrm{d}^2U/\mathrm{d}y^2$, concentrated. In smooth-wall flows, this height is roughly $y_c^+ \approx 5$--$10$. For the present cases, we observe that the instability wavelengths predicted by the stability analysis correlate well with the full shear length $L_s + y_c$, if we take $y_c^+ \approx 5$, as shown in figure~\ref{fig:instab_wavelength_corr}($a$). This suggests that the shear-layer semi-thickness above the canopies, $y_c$, is roughly constant for most of the geometries considered here, and remains close to the smooth-wall value, {while the semi-thickness below has the standard form $L_s = U/(dU/dy)$, measured at the canopy tips plane}. The only notable deviation is for the sparsest canopy studied, S48. For canopies with large element spacings, we observe that the peak in $\mathrm{d}^2U/\mathrm{d}y^2$ moves closer to the canopy-tip plane, so $y_c^+ = 5$ may no longer be a reasonable approximation for the shear-layer semi-thickness above. {Regarding the height of the shear layer within the canopy, $L_s$, we observe that it is set by the mean canopy drag coefficient, $L_s \propto \sqrt{1/C_x}$, as also noted in the studies of aquatic canopy flows by \citet{Nepf2007} and \citet{White2007}. The drag coefficient on the mean flow, in turn, depends on the element spacing and the width-to-spacing ratio}, as shown in figures~\ref{fig:instab_wavelength_corr}($b$) and ($c$). The correlation of $L_s$ with $s$, therefore, explains the dependence of the most amplified wavelength on the element spacing observed in the DNSs and the stability analysis.

\subsection{Analysis on modelled velocity profiles}
In this section we introduce a simple model for the mean velocity profile in dense canopy flows and discuss the results from their stability analysis. As we only consider canopies with small element spacings, $s^+ = \mathcal{O}(10)$, the magnitude of inertial effects within the canopies are also small and are thus neglected in the model. The results discussed in \S\ref{sec:turb_fluct} also suggest that, for very dense canopies, {turbulence and, consequently, the Reynolds shear} stresses do not penetrate within \citep{Nepf2007}, and are smooth-wall-like above the canopy-tip plane. The mean velocity above the canopy could then be modelled using a smooth-wall eddy viscosity, with the canopy-tip plane acting as the location of the smooth-wall \citep{Jimenez2001,Garcia-Mayoral2011,GG2018,GG2019}. The equation for the mean velocity can then be written as
\begin{equation}
  \frac{\mathrm{d}}{\mathrm{d}y}\left( \left[\nu + \nu_T(y)\right] \frac{\mathrm{d}U}{\mathrm{d}y}\right) - {\nu C_x(y)} U - \frac{\mathrm{d} P}{\mathrm{d} x} = 0,
\label{eq:mean_vel_model}
\end{equation}
where $C_x(y)$ is the average streamwise canopy drag coefficient, which is assumed constant within the canopy and zero outside, and $\nu_T(y)$ is the height-dependent eddy viscosity proposed by \citet{Cess1958} to approximate turbulent smooth-channel flow, and is non-zero only outside the canopy. The drag coefficients, $C_x$, used to obtain the velocity profiles are those given by equation~\eqref{eq:canopy_drag_balance} and portrayed in figure~\ref{fig:cd_vs_s}($a$). These have been obtained using the data from the direct numerical simulations, but can {also be obtained from Stokes-flow simulations as shown in figure~\ref{fig:cd_vs_s}($a$), which are significantly less computationally intensive.}

The most amplified wavelengths predicted by the stability analysis conducted on these modelled velocity profiles, with no drag applied on the fluctuations, are in reasonable agreement with those {obtained from the same no-drag analysis} on profiles obtained from the DNSs. The growth rates predicted are portrayed in figures~\ref{fig:amp_vs_lx}($g$--$i$). The wavelengths with maximum growth rates are also summarised in table~\ref{tab:instab_wavelengths}. We have also conducted stability analyses on these modelled velocity profiles including the effect of the eddy viscosity. The results are portrayed in appendix~\ref{appB}, and they are essentially the same as the ones obtained using molecular viscosity alone, apart from a slight reduction in the instability growth rates, {which suggests that although $\nu_T$ is important for setting the shape of the mean velocity profile, its effect on the fluctuations is not significant. This is likely because the Kelvin--Helmholtz-like rollers occur near the canopy tips, where $\nu_T$ is small and the molecular viscosity, $\nu$, dominates}. It is worth noting that even though this model is able to capture the instability wavelength, the velocity profiles obtained using this model do not match those from the DNSs, apart from those of S10 and G10. This is most likely due to our assumption that the turbulent stresses do not penetrate within the canopy {and remain smooth-like}, which fails as the element spacing is increased. As discussed previously, the wavelength of the instability is set by the shear length. The shear-layer semi thickness within the canopy, $L_s$, is set by the canopy drag coefficient, $C_x$. As this drag coefficient is the same both from the DNSs and for the modelled velocity profiles, we expect $L_s$ to be similar as well. The shear length above the canopy, however, could differ, as the profiles from DNS would include the effect of the turbulent stresses penetrating into the canopy and deviating from their smooth-wall values, while the modelled velocity profiles do not. The similarity in the instability wavelengths between these analyses therefore suggest that, for most of the dense canopies considered in this work, turbulence is essentially precluded from penetrating into the canopy, and that the shear length above does not vary significantly from its smooth-wall value.

\section{Conclusions}\label{sec:conclusions}
In the present work, we have examined the effect of the canopy layout on turbulent flows over canopies of densely packed filaments of small size. Three families of simulations have been conducted, the first with the element height in friction units fixed, the second with the element spacing fixed, and the third with the height-to-spacing ratio fixed. {The layouts considered had height-to-spacing ratios greater than one, and elements spacings in the range $s^+ \approx 3$--$50$. The penetration of turbulent fluctuations within such canopies was limited by their small element spacings. Consequently, the height of the roughness sublayer was also determined by the element spacing, rather than their height, extending up to $y \approx 2$--$3s$ above the canopy tips. The canopy drag coefficient was also found to be determined by the element spacing, $C_x \sim 1/s^2$. Canopies with small spacings were, therefore, found to suppress the velocity fluctuations within them owing to the large drag exerted, and the fluctuations became more intense as the spacing increased. The intensity of the characteristic Kelvin--Helmholtz-like instability over canopies was observed to be governed by two competing effects resulting from the canopy drag, the inflection at the canopy tips and the drag on the fluctuations. Canopies with large drag had a large shear at the canopy tips, and thus a stronger inflection, which enhanced the instability, but also exerted a large drag on the velocity fluctuations, which suppressed the instability. The instability was found to be inhibited in canopies with $s^+ \lesssim 10$ and, for the range of canopy spacings considered here, a stronger signature of the instability was observed as the spacing was increased. We also showed that the main contribution to the velocity fluctuations deep within the canopy was the footprint of the Kelvin--Helmholtz-like instability, and that the contribution of the element-induced dispersive flow was negligible. Short canopies with $h/s \sim 1$ were also found to inhibit the instability, owing to the blocking effect of the wall at the canopy base. For height to spacing ratios $h/s \gtrsim 6$, the instability was no longer influenced by the bottom wall, and the effect of the canopy height on the flow within and above the canopy saturated. Increasing the canopy height for a fixed spacing did not change the element-induced velocity fluctuations, and instead affected the surrounding flow through the influence of height on the instability.}

{Linear stability analysis conducted on the mean velocity profiles obtained from the DNSs is} able to capture the approximate wavelength of the instability observed in the DNSs for canopies with small element spacings. The analysis fails for larger element spacings, for which the assumption of the flow perceiving the canopy in a homogenised fashion breaks down. We showed that the shear-layer thickness, which determines the instability wavelength, {has two components, one within the canopy and the other above. The latter is set by the height above the canopy tips at which the vorticity gradient concentrates, and is essentially constant for the present canopies, $y_c^+ \approx 5$. The shear layer thickness within the canopy follows the conventional definition, $L_s = U/(dU/dy)$, and is determined by the canopy drag, thus depending linearly on the canopy spacing. We have also proposed a simplified model to capture the most amplified instability wavelength over dense canopies.} The model assumes that the turbulence above the canopy does not penetrate within and remains smooth-wall-like, and uses the mean streamwise drag coefficient of the canopies to synthesise an approximate mean-flow profile. The stability analysis conducted using these synthesised profiles yields similar results to those conducted using the mean profiles from DNS.

\bigskip
A.S. was supported by an award from the Cambridge Commonwealth, European and International Trust. Computational resources were provided by the	 ``Cambridge Service for Data Driven Discovery" operated by the University of Cambridge Research Computing Service and funded by EPSRC Tier-2 grant EP/P020259/1. The authors report no conflicts of interest.

\bigskip

\appendix

\section{{Immersed boundary method and validation}}\label{appA}
{For the present work, a modified version of the immersed boundary algorithm proposed by \citet{Garcia-Mayoral2011} is used. The algorithm of \citet{Garcia-Mayoral2011} was based on a direct-forcing approach, which applies a body force within the immersed-boundary points to drive the velocity at these points to zero \citep{Mittal2005}. The condition to implement at the points within the canopy elements is 
\begin{equation}
\frac{\mathbf{u}^{n+1} - \mathbf{u}^n}{\Delta t} = \frac{-\mathbf{u}^n}{\Delta t}
\end{equation}
Following \citet{Garcia-Mayoral2011}, this condition can be approximated by modifying the right-hand-side of equation~\eqref{eq:disc_NS}
\begin{equation}
\label{eq:IB_RGM}
  \left[\mathrm{I} - \Delta t \frac{\beta_k}{\Rey} \mathrm{L}\right] \mathbf{u}^n_{k}  =
      - \Delta t \frac{\beta_k}{\Rey} \mathrm{L} \mathbf{u}^n_{k-1}.
\end{equation}}
 \begin{figure}
	\centering
	 \includegraphics[scale=0.58]{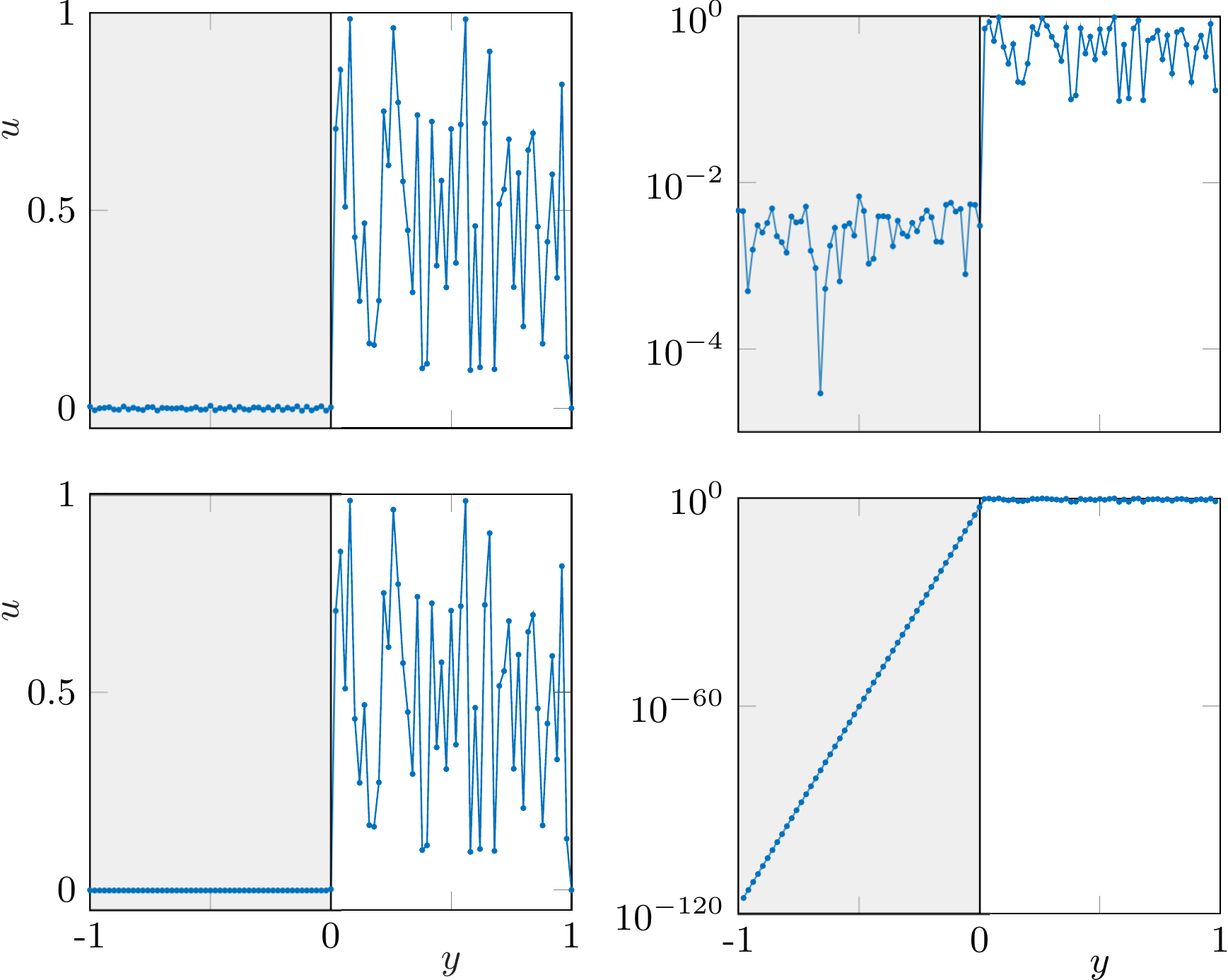}
	    \mylab{-9.5cm}{7.7cm}{($a$)}%
  		\mylab{-4.1cm}{7.7cm}{($b$)}%
  		\mylab{-9.5cm}{3.65cm}{($c$)}%
  		\mylab{-4.1cm}{3.65cm}{($d$)}%
%         \subfloat{%
%  		 \tikzsetnextfilename{IB2_lin}
%  		\input{images_val/IB_1D2_lin.tex}
%  		}%
%  		\subfloat{%
%  		 \hspace{1cm}\tikzsetnextfilename{IB2_log}
%  		\input{images_val/IB_1D2_log.tex}
%  		}%  		
%  		
%		\vspace{-0.5cm}\subfloat{%
%  		\tikzsetnextfilename{IB1_lin}
%  		\input{images_val/IB_1D1_lin.tex}
%  		}%
%  		\subfloat{%
%  		\hspace{1cm}\tikzsetnextfilename{IB1_log}
%  		\input{images_val/IB_1D1_log.tex}
%  		}%
  		 \caption{{Velocity profiles obtained using the two immersed boundary algorithms described in appendix~\ref{appA}, after one time step, starting from random initial conditions. ($a$,$b$) show results obtained from algorithm utilised by \citet{Garcia-Mayoral2011} and \citet{Abderrahaman2019}, given by equation~\eqref{eq:IB_RGM}, and ($c$,$d$) those from the present algorithm, using equation~\eqref{eq:IB_conditions}. The shaded regions mark the location of the solid obstacles. The same data is plotted in the left and right columns, except that the right column portrays the velocities in a logarithmic scale.}}
  		 	\label{fig:IB_1D}	
\end{figure} 
{The original code of \citet{Garcia-Mayoral2011} used a collocated grid for the wall-normal coordinate and was extended by \cite{Fairhall2018} to employ a staggered grid. \citet{Fairhall2018} also split the Laplacian operator on the left-hand-side of equation~\eqref{eq:disc_NS} into its wall-parallel and wall-normal components following \citet{Kim1985}}
{\begin{equation}
 \left[\mathrm{I} - \Delta t \frac{\beta_k}{\Rey} \mathrm{L}\right] \mathbf{u}  \approx  \left[\mathrm{I} - \Delta t \frac{\beta_k}{\Rey} \mathrm{L}_{xz}\right]\left[\mathrm{I}- \Delta t \frac{\beta_k}{\Rey} \mathrm{L}_{y}\right] \mathbf{u},
\end{equation}
where $L_{xz}$ includes the wall-parallel components of $L$, and $L_{y}$ the wall-normal one. Splitting the Laplacian in this manner still retains the second-order temporal accuracy of the code \citep{Kim1985}. Equation~\eqref{eq:disc_NS} can then be written as 
\begin{equation}
\left[\mathrm{I} - \Delta t \frac{\beta_k}{\Rey} \mathrm{L}_{y}\right] \mathbf{u}\  =  \left[\mathrm{I} - \Delta t \frac{\beta_k}{\Rey} \mathrm{L}_{xz}\right]^{-1}\mathrm{RHS}.\label{eq:disc_NS_RHS_mod}
\end{equation}}
{In the present work, we implement a modified version of the immersed boundary algorithm used by \citet{Garcia-Mayoral2011} into the above algorithm, which offers an improvement in the accuracy for the velocities within the immersed boundary regions. This implementation is summarised below. The right-hand-side of equation~\eqref{eq:disc_NS_RHS_mod} is then transformed to physical space, and modified to satisfy the following conditions within the immersed boundary points
\begin{equation}
\label{eq:IB_conditions}
  \left[\mathrm{I} - \Delta t \frac{\beta_k}{\Rey} \mathrm{L}_{y}\right] \mathbf{u}^n_{k}  = \left\{
    \begin{array}{ll}
      - \Delta t \frac{\beta_k}{\Rey} \mathrm{L}_{y} \mathbf{u}^n_{k-1}, & y = y_{tips} \\[2pt]
      0, & y \neq y_{tips}.
  \end{array} \right.
\end{equation}
where $y_{tips}$ denotes the wall-normal plane at the canopy tips. 
At the interfaces, the condition imposed by equation~\eqref{eq:IB_conditions} yields $\mathbf{u}^n_k  = \mathcal{O}(\Delta t^2)$, which is of the order of the temporal discretisation error of the code. As a staggered grid is used, the wall-normal grid points for the streamwise velocities are offset by half a grid spacing from those of the wall-normal velocity. The element tips are aligned with the grid for the streamwise velocity. For the wall-normal velocity, the interface condition is set at the grid point just below the canopy-tip plane, which enforces near-zero wall-normal velocity at the canopy tips through continuity. Away from the interfaces, within the immersed boundary region, the condition set by equation~\eqref{eq:IB_conditions} results in an exponential decay of the velocity in the wall-normal direction from its $\mathcal{O}(\Delta t^2)$ value at the interface. To illustrate this decay, the results from a simple, one-dimensional implementation of this algorithm are shown in figure~\ref{fig:IB_1D}. The velocity is assumed to vary only in the wall-normal direction. For these test cases, random initial conditions are used to mimic turbulent fluctuations in the flow. The velocity fields obtained for these cases, after one timestep, using the immersed boundary conditions of equation~\eqref{eq:IB_conditions} are compared to those obtained using equation~\eqref{eq:IB_RGM} in figure~\ref{fig:IB_1D}.}
\begin{figure}
        \centering
       \vspace{5mm} \subfloat{%
                   \includegraphics[scale=0.79,trim={0mm 0mm 12mm 0mm},clip]{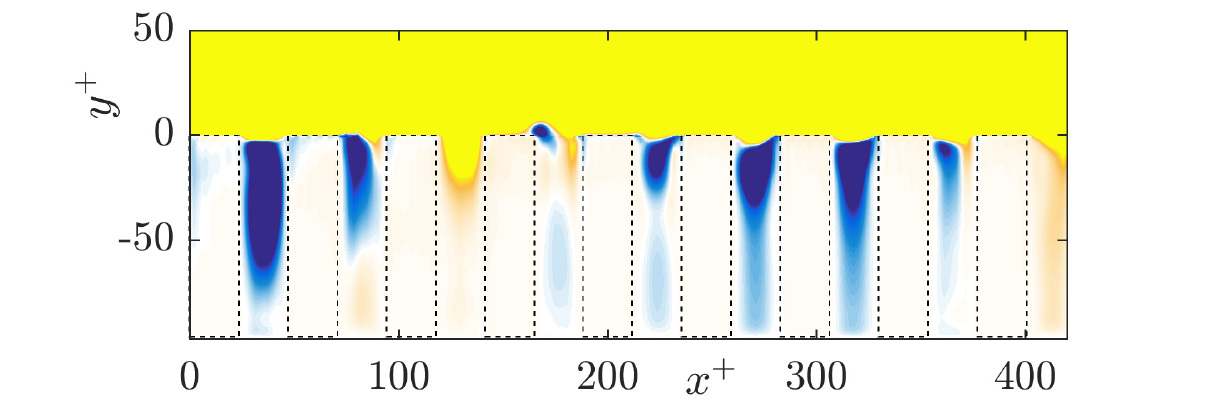}%
         }%
         
         \vspace{-8mm} \subfloat{%
                   \includegraphics[scale=0.79,trim={0mm 0mm 12mm 0mm},clip]{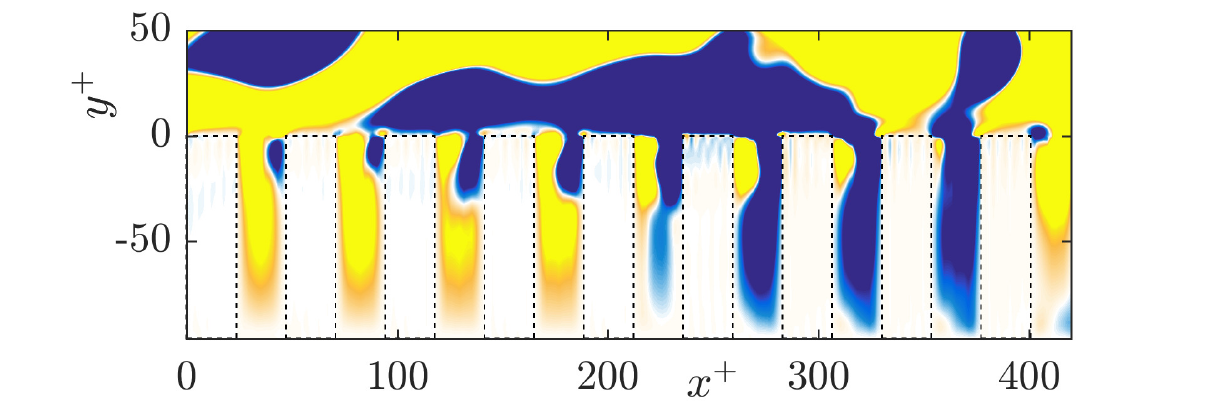}%
         }%
        
        \vspace{-8mm} \subfloat{%
                   \includegraphics[scale=0.8,trim={0mm 0mm 12mm 0mm},clip]{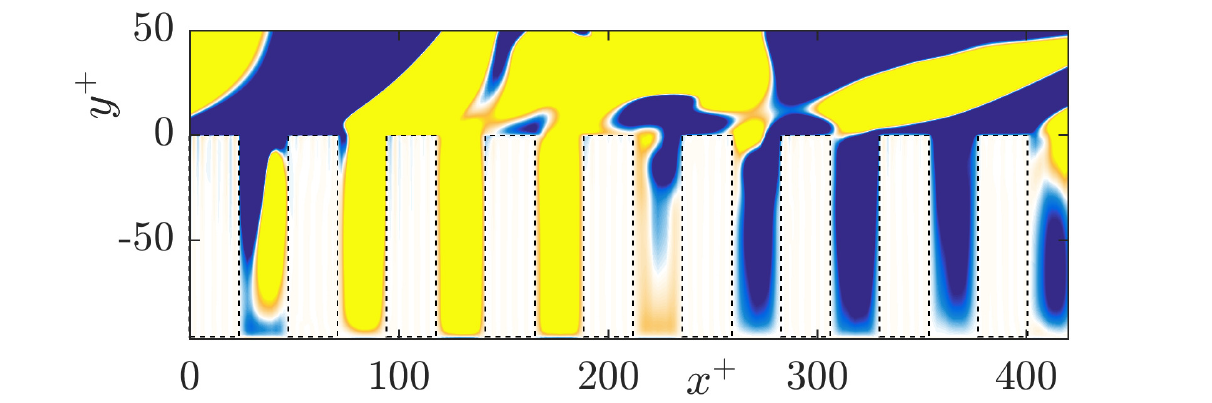}%
         }%
         \mylab{-8.5cm}{5.4cm}{($a$)}%
        \mylab{-8.5cm}{2.6cm}{($b$)}%
        \mylab{-8.5cm}{-0.2cm}{($c$)}%
          \caption{{Instantaneous realisations of the ($a$) streamwise, ($b$) wall-normal and ($c$) spanwise velocities in a plane passing through the middle of the canopy elements for case S48, scaled with the friction velocity $u_\tau$. The clearest and darkest contours represent intensities of $\pm 0.1$, respectively.}}%
            \label{fig:val_snaps}
\end{figure} 
\begin{figure}
	\centering
		\subfloat{%
%  		\tikzsetnextfilename{grid}
  		\includegraphics[scale=1]{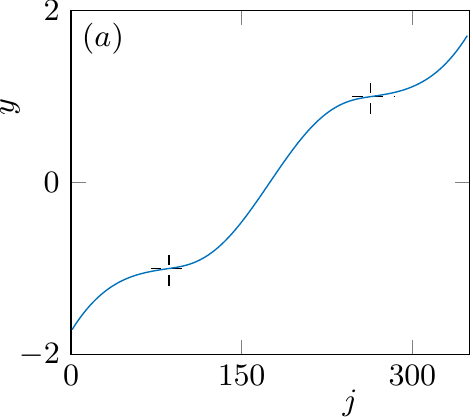}
  		}%
  		\subfloat{%
%  		 \tikzsetnextfilename{dgrid}
  		\hspace{4mm}\includegraphics[scale=1]{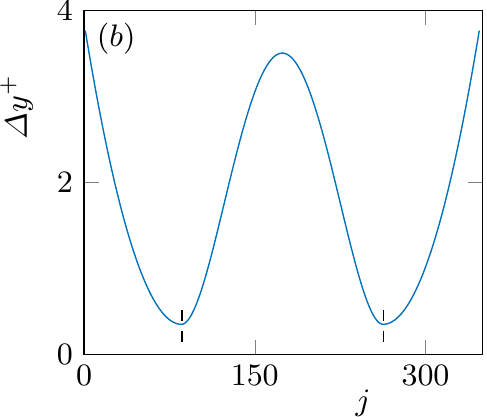}
  		}%
  		 \caption{{Wall-normal grid distribution for the simulation of case S10. ($a$) Variation of the wall-normal coordinate and ($b$) grid resolution with an equispaced auxiliary variable $j$. Dashed lines mark the location of the canopy tip plane.}}
 		  	\label{fig:grid}	
\end{figure}    
 \begin{figure}
	\centering
		\subfloat{%
%  		\tikzsetnextfilename{u_val}
  		\hspace{0mm}\includegraphics[scale=1]{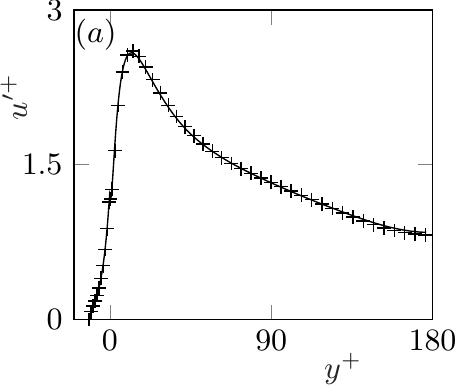}
  		}%
  		\subfloat{%
%  		\tikzsetnextfilename{v_val}
  		\hspace{-1mm}\includegraphics[scale=1]{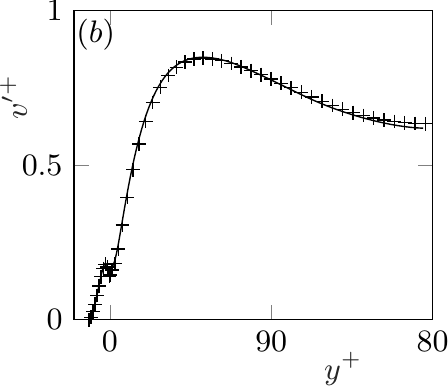}
  		}%
  		\subfloat{%
%  	 \tikzsetnextfilename{w_val}
  		\hspace{0mm}\includegraphics[scale=1]{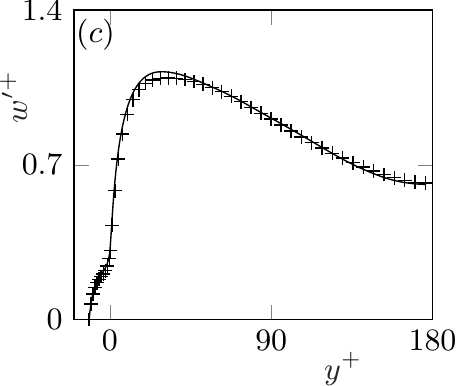}
  		}%
  		
  		\vspace{0mm}\subfloat{%
%  		 \tikzsetnextfilename{Umean_val}
  		\hspace{0mm}\includegraphics[scale=1]{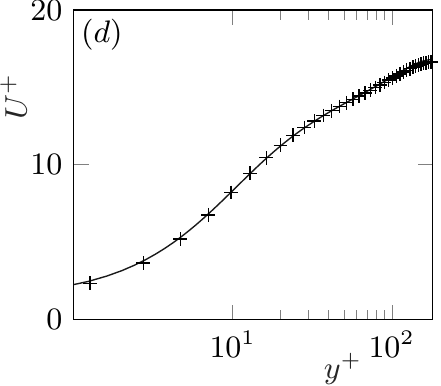}
  		}%
  		\subfloat{%
%  		 \tikzsetnextfilename{uv_val}
  		\hspace{0mm}\includegraphics[scale=1]{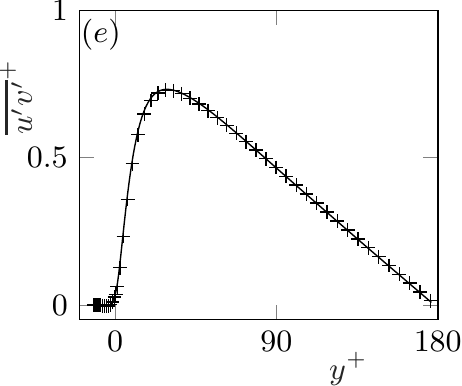}
  		}%
  		 \caption{{Rms velocity fluctuations, mean velocity and Reynolds shear stress profiles. The
solid lines represent the results obtained from the present code, and the $+$ symbol represent
the data of case C12 from \citet{Abderrahaman2019}.}}
 		  	\label{fig:stats_val}	
\end{figure}

 \begin{figure}
	\centering
		\vspace{0mm}\subfloat{%
%  		 \tikzsetnextfilename{u_t9-18-27_grid_ind}
  		\hspace{0mm}\includegraphics[scale=1,trim=0mm 0mm 0mm 0mm,clip]{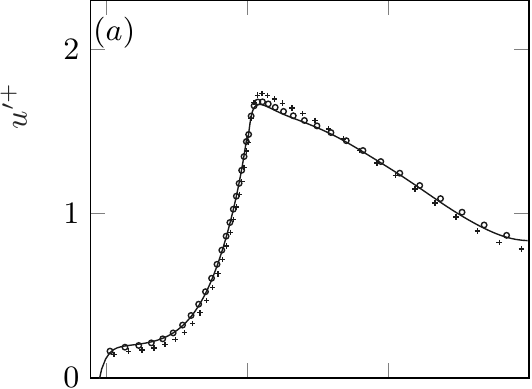}
  		}%
  	   \subfloat{%
%  		 \tikzsetnextfilename{u_t12-24-36_grid_ind}
  		\hspace{4mm}\includegraphics[scale=1,trim=0mm 0mm 0mm 0mm,clip]{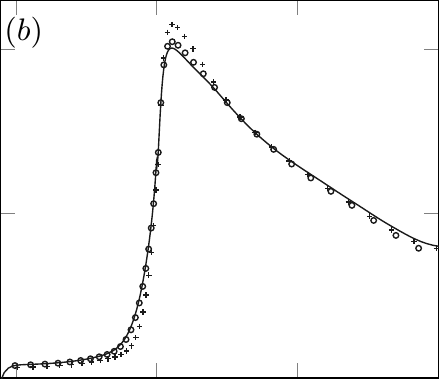}
  		}%
  		
  		\vspace{0mm}\subfloat{%
%  		 \tikzsetnextfilename{v_t9-18-27_grid_ind}
  			\hspace{0mm}\includegraphics[scale=1,trim=0mm 0mm 0mm 0mm,clip]{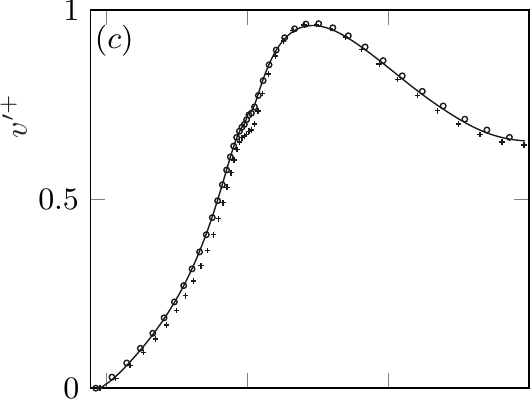}
  		}%
  		\subfloat{%
%  		 \tikzsetnextfilename{v_t12-24-36_grid_ind}
  			\hspace{4mm}\includegraphics[scale=1,trim=0mm 0mm 0mm -0.7mm,clip]{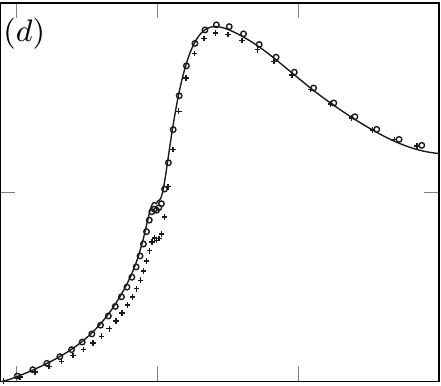}
  		}%
  		
  		\vspace{0mm}\subfloat{%
% 		\tikzsetnextfilename{w_t9-18-27_grid_ind}
 			\hspace{0mm}\includegraphics[scale=1,trim=0mm 0mm 0mm 0mm,clip]{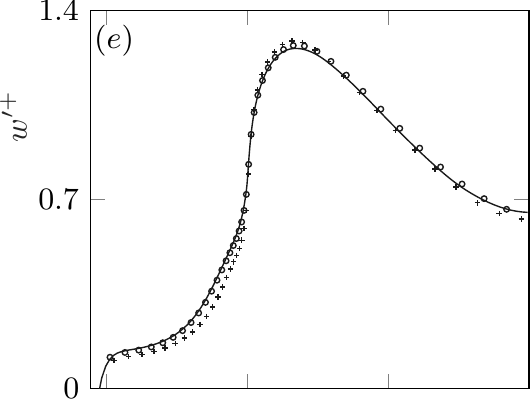}
 		}%
 		\subfloat{%
% 		 \tikzsetnextfilename{w_t12-24-36_grid_ind}
 			\hspace{4mm}\includegraphics[scale=1,trim=0mm 0mm 0mm -0.7mm,clip]{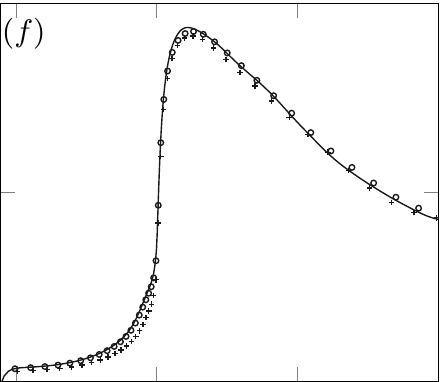}
 		}%
 		
 		\vspace{0mm}\subfloat{%
% 		\tikzsetnextfilename{uv_t9-18-27_grid_ind}
 		\hspace{1.7mm}\includegraphics[scale=1,trim=0mm 0mm 0mm 0mm,clip]{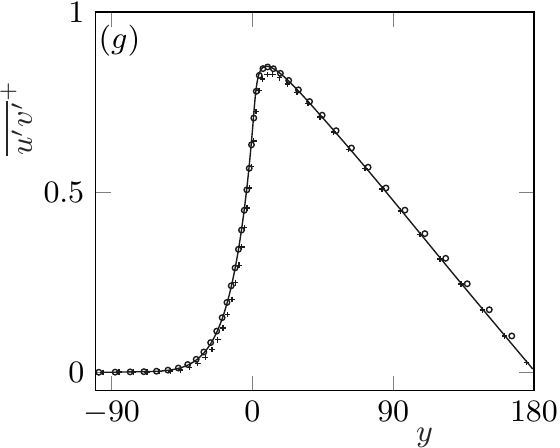}
 		}%
 		\subfloat{%
% 		\tikzsetnextfilename{uv_t12-24-36_grid_ind}
 		\hspace{0.6mm}\includegraphics[scale=1,trim=0mm 0mm 0mm -1.05mm,clip]{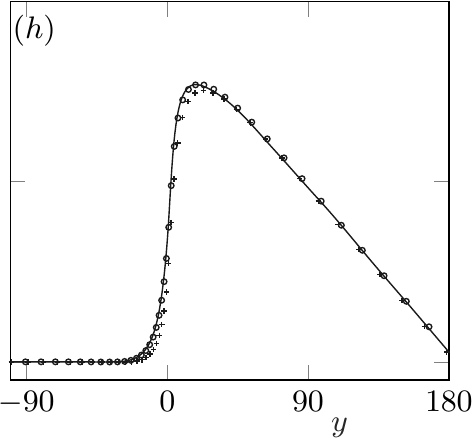}
 		}%
 		 \caption{Rms velocity fluctuations and Reynolds shear stress profiles. The panels first column represent case G100 and those in the second column represent case S48. The solid lines in ($a$,$c$,$e$,$g$) represent the results from using $27$ points per spacing; the symbols \protect\hcircle, $18$ points; and $+$, $9$ points per spacing. The solid lines in ($b$,$d$,$f$,$h$) represent the results from using $36$ points per spacing; the symbols \protect\hcircle, $24$ points; and $+$, $12$ points per spacing.}
 		  	\label{fig:stats_grid_ind}	
\end{figure}	

{Although both algorithms result in small velocities within the solid regions, the implementation proposed here results in a smoother and much faster decay of the velocity within the solid. In the DNS code, however, the velocity correction step introduces an error of order $\Delta t^2$ in all the immersed boundary points. Even so, in experience it was observed that the proposed algorithm is a more stable numerical implementation of the immersed boundaries compared to the one proposed by \citet{Garcia-Mayoral2011}. This is likely due to the present method not generating sharp gradients in the velocity field within the solid obstacles at the pressure calculation step. The velocity within the canopy elements, or the permeability error, in the DNSs is observed to be less than $0.1u_\tau$, for all the conducted simulations, and is much smaller than velocity in the `fluid' points surrounding the elements. This is illustrated in the instantaneous realisations of the velocity fields from one of the simulations are portrayed in figure~\ref{fig:val_snaps}. For completeness, the wall-normal grid distribution for case S10 is portrayed in figure~\ref{fig:grid}. In order to validate the implementation of the immersed boundaries, we have replicated the DNS for the collocated roughness elements with height $h^+ \approx 12$ of \citet{Abderrahaman2019}. The mean velocity profiles, rms fluctuations and the Reynolds shear stresses obtained from the present simulations show good agreement with the results of \citet{Abderrahaman2019}, as shown in figure~\ref{fig:stats_val}. We have also conducted a grid-dependence analysis, for which the results are portrayed in figure~\ref{fig:stats_grid_ind}. Velocity rms fluctuations for cases S48 and G100 are shown for different wall-parallel resolutions.}

 \begin{figure}
	\centering
		\subfloat{%
%  		 \tikzsetnextfilename{omega_lx_S_comp}
  		\hspace{0mm}\includegraphics[scale=1]{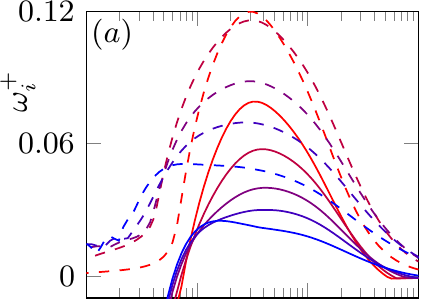}
  		}%
  		\subfloat{%
%  		 \tikzsetnextfilename{omega_lx_H_comp}
  		\hspace{1mm}\includegraphics[scale=1]{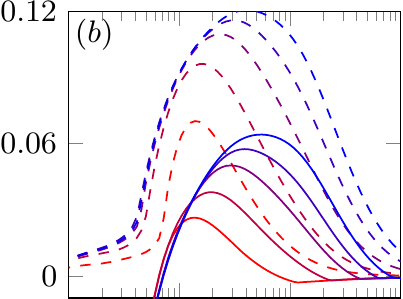}
  		}%
  		\subfloat{%
%  		 \tikzsetnextfilename{omega_lx_G_comp}
  		\hspace{1mm}\includegraphics[scale=1]{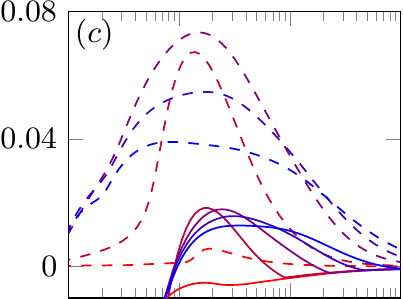}
  		}%
  		
  	\vspace{0mm}\subfloat{%
%  		 \tikzsetnextfilename{omega_lx_S_z_comp}
  		\hspace{0mm}\includegraphics[scale=1]{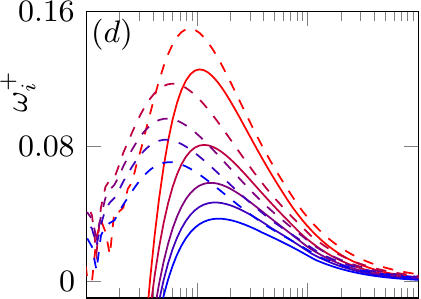}
  		}%
  	     \subfloat{%
%  		 \tikzsetnextfilename{omega_lx_H_z_comp}
  		\hspace{1mm}\includegraphics[scale=1]{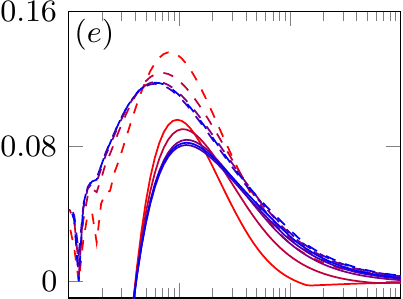}
  		}%
  		\subfloat{%
%  		 \tikzsetnextfilename{omega_lx_G_z_comp}
  		\hspace{1mm}\includegraphics[scale=1]{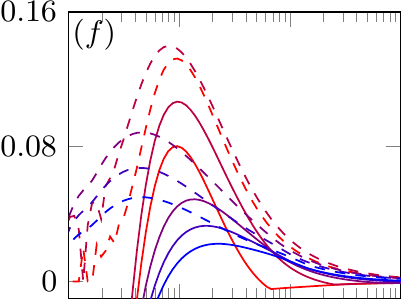}
  		}%
  		
  		 \vspace{0mm}\subfloat{%
%  		 \tikzsetnextfilename{omega_lx_m_S_z_comp}
  		\hspace{1.8mm}\includegraphics[scale=1]{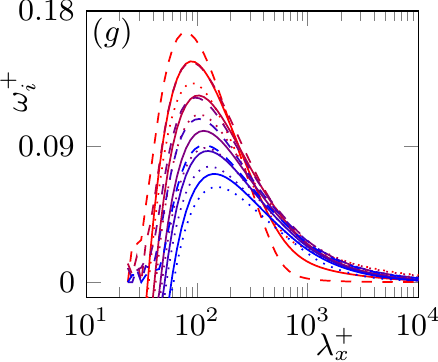}
  		}%
  	     \subfloat{%
%  		 \tikzsetnextfilename{omega_lx_m_H_z_comp}
  		\hspace{-0.8mm}\includegraphics[scale=1]{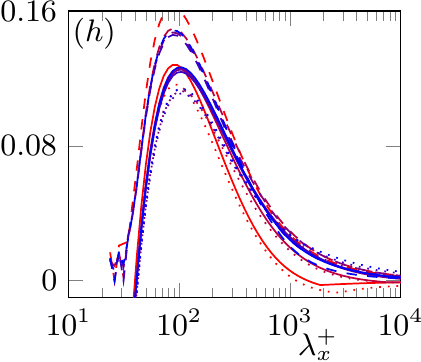}
  		}%
  		\subfloat{%
%  		 \tikzsetnextfilename{omega_lx_m_G_z_comp}
  		\hspace{-0.8mm}\includegraphics[scale=1]{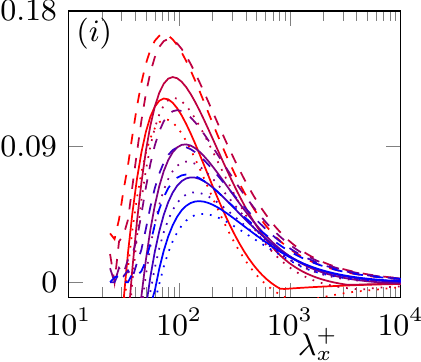}
  		}%
			 \caption{Growth rates of different perturbation wavelengths obtained from the stability analysis performed on ($a$--$c$) mean profiles obtained from the DNSs, with drag on the perturbations included in the stability analysis; ($d$--$f$) mean profiles obtained from DNSs, with no drag on the perturbations; and ($g$--$i$) mean velocity profiles obtained using equation~\eqref{eq:mean_vel_model}, with no drag on the perturbations. \protect\blackline, viscous analysis including molecular viscosity alone; \protect\blackdashed, inviscid analysis; \protect\blackdotted, viscous analysis including an eddy viscosity. The colours from red to blue represent ($a$,$d$,$g$) cases S10 to S48; ($b$,$e$,$h$) cases H16 to H128; and ($c$,$f$,$i$) cases G10 to G100.}
 		  	\label{fig:amp_vs_lx_comp}	
\end{figure}

\section{Comparison of inviscid and viscous stability analysis}\label{appB}
Here, we provide the governing equations used to perform a stability analysis with a turbulent viscosity varying in the wall-normal direction, $\nu_T(y)$. The modified Orr-Sommerfeld equation is then given by
\begin{eqnarray}
(\alpha U - i C_y)&&(\mathrm{D}^2   -  \alpha^2)\widetilde{v}
- \alpha U'' \widetilde{v}  - i(C_x - C_y) D^2 \widetilde{v} + 2i \nu_T'(\mathrm{D}^3 - \alpha^2 \mathrm{D})\widetilde{v}  \\  && + {i\nu_T}(\mathrm{D}^4 - 2\alpha^2 \mathrm{D}^2 + \alpha^4)\widetilde{v}+ i \nu_T''(\mathrm{D}^2 + \alpha^2)\widetilde{v}= \omega (\mathrm{D}^2 - \alpha^2) \widetilde{v}.
\label{eq:OS_app_B}
\end{eqnarray}
A similar equation, excluding the canopy drag terms, has also been used by \citet{Reynolds1972}, \citet{DelAlamo2006}, \citet{Pujals2009} and \cite{GG2018}. In figures~\ref{fig:amp_vs_lx_comp}($a$--$f$), we compare the results obtained from viscous and inviscid analysis conducted using the velocity profiles from the DNSs. In In figures~\ref{fig:amp_vs_lx_comp}($g$--$i$), we show the results from inviscid and viscous stability analyses, using both molecular and turbulent viscosities, performed using the velocity profiles obtained from equation~\eqref{eq:mean_vel_model}.

\bibliographystyle{jfm}
\bibliography{jfm_ref.bib}

\end{document}